\documentclass[onecollarge]{svjour2}       
\usepackage{graphicx}
\begin{document}

\title{Tunneling time problem: At the intersection of quantum mechanics, classical probability theory and special relativity}


\author{Nikolay L. Chuprikov 
}


\institute{N. L. Chuprikov \at
              Tomsk State Pedagogical University, 634041, Tomsk, Russia \\
              \email{chnl@tspu.edu.ru}           
}

\date{Received: date / Accepted: date}

\maketitle
\begin{abstract}
After the review by Hauge and Stovneng the old question of "How long does it take to tunnel through the barrier?" has not still lost its
relevance. As before, there is no clear answer to this question even for the one-dimensional completed scattering (OCS). In this paper we show
that this seemingly simple question stands alongside with such fundamental problems of quantum mechanics (QM) as the Schrodinger's-cat and and
EPR-Bohm paradoxes. Their common feature is that the states of a scattered particle, a radioactive atom and electron EPR-pair represent pure {\it
micro-cat} states. It is widely accepted that the EPR-Bohm paradox implies the non-existence of local hidden variables (LHVs), the cat paradox
represents a macro-objectification problem and the tunneling time is an observer-dependent quantity whose definitions must be 'operational'. At
the same time, according to the probabilistic approach (Accardi, Khrennikov, Philipp, Hess {\it et al.}) to Bell's inequality that underlies the
EPR-Bohm experiments, its experimental violation means simply that it contains probability distributions associated with mutually incompatible
statistical data. We argue that this approach must be extended onto micro-cat states because they describe, too, mutually incompatible statistical
data. The current practice to interpret the squared modulus of a micro-cat state as the probability density should be recognized as erroneous. It
is this practice that makes QM incompatible with classical physical theories and, thereby, makes the micro-world 'unspeakable'. The known
'operational' tunneling-time concepts, elaborated in line with this practice, are logically inconsistent. We argue that the TTP and the cat
paradox must be solved at the level of single electrons and atoms, without resorting to environment and measurement contexts. We present a new
model of the OCS and solve the TTP on its basis.

\end{abstract}

\newcommand{\Api}{A^{in}}
\newcommand{\Ami}{B^{in}}
\newcommand{\Apo}{A^{out}}
\newcommand{\Amo}{B^{out}}
\newcommand {\uta} {\tau_{tr}}
\newcommand {\utb} {\tau_{ref}}
\newcommand{\ppp}{\mbox{\hspace{5mm}}}
\newcommand{\ooo}{\mbox{\hspace{3mm}}}
\newcommand{\ooa}{\mbox{\hspace{1mm}}}
\newcommand{\ppa}{\mbox{\hspace{25mm}}}
\newcommand{\ppb}{\mbox{\hspace{35mm}}}
\newcommand{\ppc}{\mbox{\hspace{10mm}}}

\section{Introduction} \label{intro}

The TTP, with its key question of "How long does it take to tunnel through the barrier?", is one of long-standing and controversial problems in
QM. The modern tunneling time literature (TTL) (see, e.g., reviews \cite{Ha2,La1,Olk1,Nus,Mug,Ste,Win} as well as original papers
\cite{Lun,SoAh,Gros,Xue} published after the last journal review \cite{Win}) contains a huge variety of contenders on the role of the tunneling
time, but none of them gives a flawless answer to this question. As was said in \cite{Ha2}, "All [the known concepts] have been found to suffer
one logical flaw or another, flaws sufficiently serious that must be rejected" (see also \cite{Nus,Win}).

At the same time most researchers dealing with the TTP are inclined to believe that this problem has already been solved, that all the existing
'operational' tunneling-time concepts give correct answers to the key question for relevant measurement contexts (see, e.g.,
\cite{La1,Olk1,Mug,Ste}). Moreover, it is widely believed that one of these concepts -- the Wigner time -- has already been measured in the
single-photon experiments \cite{Ste2,Pol}, and the Hartman effect predicted on its basis has been observed.

For the proponents of this viewpoint the main problem is to reconcile the superluminal velocities observed in these experiments with special
relativity (SR). As regards the logical flaws of the Wigner time concept, they are considered to be unimportant as compared with the fact that
this quantity has been measured. As is believed, this concept is inconsistent only from the viewpoint of {\it classical} physical theories which
imply the existence of LHVs. But Bell's theory they claim rejects the existence of LHVs and hence the classical logics is inapplicable to the
tunneling time which does not exist as a LHV.

But is it right? Is this approach to the TTP, based on Bell's theory of LHVs, internally consistent? To what extent is justified Bell's inference
on the nonexistence of LHVs? Why the probabilistic interpretation of the Bell inequalities is not taken into account when solving the TTP?

To answer these questions and to present a new way of solving the TTP is the goal of our paper. Its plan is as follows: to review the most
prominent 'operational' approaches to the TTP (Section \ref{TTL}) and most prominent explanations of the Hartman effect (Section \ref{hart}); to
dwell separately on the peculiarities of the Bohmian model of tunneling (Section \ref{Bohm}); to analyse the most prominent modern schemes of
measuring the tunneling time in QM (Sections \ref{weak} and \ref{experiment}); to show that the non-existence of LHVs has been postulated, in
fact, rather than proven (Section \ref{req}); to present a new model of the OCS (Section \ref{separ}) and results of studying the temporal aspects
of tunneling obtained on its basis (Sections \ref{f2} and \ref{hart1}).

\section{Conventional quantum-mechanical model of a one-dimensional completed scattering} \label{smodel}

We begin with the conventional model of scattering a particle on a one-dimensional static (non-oscillating) potential barrier. For definiteness,
we shall consider the case when a particle impinges from the left at the symmetric one-dimensional potential barrier $V(x)$ situated in the
spatial interval $[a,b]$: $V(x_c-x)=V(x-x_c)$ where $x_c=(b+a)/2$ is the midpoint of the barrier region; $a>0$.

For the incident particle with a fixed momentum $\hbar k$ the wave function $\Psi_{tot}(x;k)$ is
\begin{eqnarray} \label{1}
\Psi_{tot}(x;k)=\left\{ \begin{array}{c} e^{ikx}+b_{out}(k)e^{ik(2a-x)}\ppa x< a\\
a_{tot} f(x-x_c;k)+b_{tot} g(x-x_c;k) \ppp a< x< b\\
a_{out}(k)e^{ik(x-d)}\ppb x>b
\end{array}\right. ;
\end{eqnarray}
$d=b-a$ is the barrier width; $f(x-x_c;k)$ and $g(x-x_c;k)$ are such real independent partial solutions to the Schr\"odinger equation in the
interval $[a,b]$ that $f(x_c-x;k)=-f(x-x_c;k)$, $g(x_c-x;k)=g(x-x_c;k)$ and $f_x g-g_x f=\kappa$ where $\kappa$ is a positive constant;
$f_x=df(x-x_c;k)/dx$, $g_x=dg(x-x_c;k)/dx$.

For example, if $V(x)$ is the rectangular barrier of height $V_0$ and $E<V_0$ ($E=\hbar^2k^2/2m$), then
\begin{eqnarray*}
f=\sinh[\kappa (x-x_c)],\ooo g=\cosh[\kappa(x-x_c)],\ooo \kappa=\sqrt{2m(V_0-E)}/\hbar.
\end{eqnarray*}

Since $\Psi_{tot}(x;k)$ and its first spatial derivative must be continuous at the points $x=a$ and $x=b$,
\begin{eqnarray*}
a_{tot}=-\frac{e^{ika}}{\kappa}\ooa a_{out}P^*=\frac{e^{ika}}{\kappa}\left(P+b_{out}P^*\right),\ppp P=g_x(b-x_c;k)+ikg(b-x_c;k);\\
b_{tot}=\frac{e^{ika}}{\kappa}\ooa a_{out}Q^*=\frac{e^{ika}}{\kappa}\left(Q+b_{out}Q^*\right),\ppp Q=f_x(b-x_c;k)+ikf(b-x_c;k).
\end{eqnarray*}
From here it follows that
\begin{eqnarray} \label{2}
a_{out}=\frac{1}{2}\left(\frac{Q}{Q^*}-\frac{P}{P^*}\right),\ooo b_{out}= -\frac{1}{2}\left(\frac{Q}{Q^*}+\frac{P}{P^*}\right).
\end{eqnarray}

These two quantities can be written also via the elements of the transfer matrix \textbf{Y}:
\begin{eqnarray*}
\left(
\begin{array}{rl}
1 \\ b_{out}e^{2ika}
\end{array} \right)=\textbf{Y}
\left(
\begin{array}{rl}
a_{out}e^{-ikd} \\ 0
\end{array} \right);\ppp \textbf{Y} =\left(
\begin{array}{rl}
q & p \\ p^* & q^*
\end{array} \right).
\end{eqnarray*}
For any barrier in the interval $[a,b]$ the elements $q$ and $p$ can always be written in the form (see \cite{Ch8})
\begin{eqnarray*}
q=\frac{1}{\sqrt{T}}e^{i[k(b-a)-J]},\ppp p=i\sqrt{\frac{R}{T}}\ooa e^{i[F-k(b+a)]}
\end{eqnarray*}
where $J(k)$ and $F(k)$ are two phases, $T(k)$ and $R(k)$ are (real) transmission and reflection coefficients, respectively; $T(k)+R(k)=1$. Thus,
\begin{eqnarray} \label{3}
a_{out}=\sqrt{T}\ooa e^{iJ},\ppp b_{out}= -i\sqrt{R}\ooa e^{i(J-F)}.
\end{eqnarray}
For symmetric barriers the phase $F$ takes only two values \cite{Ch8}, either $0$ or $\pi$. From Exps. (\ref{2}) and (\ref{3}) it follows that
\begin{eqnarray} \label{3a}
\sqrt{\frac{R}{T}}=\frac{|f_x g_x+k^2 f g|}{k\kappa},\ooo J=\arctan(k^2 f g-f_x g_x, k (f_x g+f g_x));
\end{eqnarray}
$F=0$ if $f_x g_x+k^2 f g\geq 0$; otherwise, $F=\pi$ (the function $F(k)$ is discontinuous at the resonance points). In particular, for the
rectangular barrier with $E<V_0$
\begin{eqnarray} \label{3b}
\sqrt{\frac{R}{T}}=\theta_{(+)}\sinh(\kappa d),\ooo J=\arctan(\theta_{(-)}\tanh(\kappa d)),\ooo F=0;\ooo
\theta_{(\pm)}=\frac{1}{2}\left(\frac{k}{\kappa}\pm\frac{\kappa}{k}\right).
\end{eqnarray}

Now, in the time-dependent case, the particle's state can be written as
\begin{eqnarray} \label{4}
\Psi_{tot}(x,t)=\frac{1}{\sqrt{2\pi}}\int_{-\infty}^{\infty} \mathcal{A}(k)\Psi_{tot}(x;k)e^{-iE(k)t/\hbar} dk;
\end{eqnarray}
where $\mathcal{A}(k)$ is, for example, the Gaussian function: $\mathcal{A}(k)=(2l_0^2/\pi)^{1/4}\exp[-l_0^2(k-k_0)^2]$. In this setting of the
problem, the wave function at the initial time $t=0$ represents the wave packet of width $l_0$ ($l_0\ll a$), with the center of "mass" (CM)
positioned at the point $x=0$.

The incident $\Psi_{inc}(x,t)$, transmitted $\Psi_{tr}(x,t)$ and reflected $\Psi_{tr}(x,t)$ wave packets have the forms
\begin{eqnarray*}
\Psi_{inc}(x,t)=\frac{1}{\sqrt{2\pi}}\int_{-\infty}^{\infty} \mathcal{A}(k)e^{i[kx-iE(k)t]/\hbar}dk
\end{eqnarray*}
\begin{eqnarray} \label{5}
\Psi_{tr}(x,t)=\frac{1}{\sqrt{2\pi}}\int_{-\infty}^{\infty}\mathcal{A}(k)\sqrt{T(k)}e^{i\left[J(k)+k(x-d)-E(k)t/\hbar\right]}dk,
\end{eqnarray}
\begin{eqnarray*}
\Psi_{ref}(x,t)=\frac{-i}{\sqrt{2\pi}}\int_{-\infty}^{\infty}\mathcal{A}(k)\sqrt{R(k)} e^{i\left[J(k)-F(k)+k(2ka-x)-E(k)t/\hbar\right]}dk;
\end{eqnarray*}
note that $T(-k)=T(k)$, $J(-k)=-J(k)$ and $F(-k)=\pi-F(k)$.

One needs to distinguish between the following two radically different cases: the so-called one-dimensional completed scattering (OCS) and the
one-dimensional non-completed scattering (ONCS). When the scattered wave packets $\Psi_{tr}(x,t)$ and $\Psi_{ref}(x,t)$ do not overlap each other
in the limit $t\to \infty$, the time-dependent scattering process represents the OCS. This implies that the rate of scattering these packets is
much larger than the rate of widening each packet: the values of $\hbar k_0$ and $l_0$ must be large enough. Otherwise, we deal with the ONCS; in
this case the wave packets $\Psi_{tr}(x,t)$ and $\Psi_{ref}(x,t)$ overlap each other in the limit $t\to\infty$.

In the case of the OCS, the norms $\textbf{T}_{as}$ and $\textbf{R}_{as}$, where $\textbf{T}_{as}=\langle\Psi_{tr}|\Psi_{tr}\rangle$ and
$\textbf{R}_{as}=\langle\Psi_{ref}|\Psi_{ref}\rangle$, obey the "either-or" rule $\textbf{T}_{as}+\textbf{R}_{as}=1$ that reflects the principle
of additivity of probability on disjoint events. Thus, in the course of the OCS the incident wave packet $\Psi_{inc}(x,t)$ that describes the
ensemble of incident particles splits into the two disjoint wave packets $\Psi_{tr}(x,t)$ and $\Psi_{ref}(x,t)$ that describe the subensembles of
transmitted and reflected particles, respectively: each particle of the incident ensemble is {\it either} transmitted {\it or} reflected by the
barrier.

The question of how many time transmitted particles spend on average inside the barrier region is obvious to be relevant only to the OCS. In the
case of the ONCS, to speak about transmission and reflection is inappropriate.

\section{Critique of the existing concepts and timekeeping procedures in the TTL} \label{TTL}

\subsection{The dwell time} \label{dwell}

We begin our analysis with the dwell time which is considered now as a well-defined measurable (see Section \ref{PDS}) quantity that describes the
tunneling duration (see Sections \ref{larmor}). In QM, the dwell time is defined via the velocity associated with the probability flow density,
i.e., via the flow velocity.

To fix the intrinsic properties of the dwell time, we first consider the case when the interval $[a,b]$ is empty. The ensemble of free particles
is described by the wave function $\Psi_{tot}(x;k)=\Psi_{free}(x;k)=e^{ikx}$. The velocity $v_{flow}^{free}$ of particles in this ensemble is
defined via the probability flow density $I_{free}$, which can be written as $I_{free}=|\Psi_{free}(x;k)|^2 v_{flow}^{free}$. Thus, for this
velocity and the average time $\tau_{dwell}^{free}$ spent by a particle in the interval $[a,b]$, we have
\begin{equation} \label{7}
v_{flow}^{free}=\frac{I_{free}}{|\Psi_{free}(x;k)|^2}=\frac{\hbar k}{m},\ppp
\tau_{dwell}^{free}=\frac{1}{I_{free}}\int_a^b |\Psi_{free}(x;k)|^2 dx=\frac{md}{\hbar k}.
\end{equation}
As is seen from (\ref{7}), for a free particle the flow velocity $v_{flow}^{free}$ is a transit velocity, and the dwell time $\tau_{dwell}^{free}$
is thus a transit time.

When the interval $[a,b]$ is occupied by the potential barrier $V(x)$ the wave function to describe scattering a particle by this barrier is given
by Exps. (\ref{1}), and the corresponding probability flow density is $I_{tot}=I_{tr}=T(k)\hbar k/m$. Now, for the flow velocity and the dwell
time we have
\begin{equation} \label{8}
v_{flow}^{(1)}=\frac{I_{tot}}{|\Psi_{tot}(x;k)|^2},\ppp \tau_{dwell}^{(1)}=\frac{1}{I_{tot}}\int_a^b
|\Psi_{tot}(x;k)|^2 dx.
\end{equation}

An analogue of the dwell time $\tau_{dwell}^{(1)}$ is widely used (see, e.g., \cite{Ric,Mou,Spi}) in classical electrodynamics (CED). At the same
time the physical meaning of $\tau_{dwell}^{(1)}$ is unclear. The point is that unlike the wave function $\Psi_{tot}(x;k)$ the flow density
$I_{tot}(x;k)$ relates only to transmitted particles. Due to this peculiarity of $\Psi_{tot}(x;k)$ (see also Sections \ref{Bohm} and
\ref{conclude}) the quantities $v_{flow}^{(1)}$ and $\tau_{dwell}^{(1)}$ characterize neither the whole ensemble of scattering particles nor its
transmitted part; $v_{flow}^{(1)}$ has no relation to the velocity of transmitted particles in the barrier region.

In QM, of popular is Buttiker's version of the dwell time \cite{But} (see also Section \ref{larmor}):
\begin{equation} \label{9}
v_{flow}^{(2)}=\frac{I_{inc}}{|\Psi_{tot}(x;k)|^2},\ppp \tau_{dwell}^{(2)}=\frac{1}{I_{inc}}\int_a^b
|\Psi_{tot}(x;k)|^2 dx,
\end{equation}
where $I_{inc}=\hbar k/m$. The probability flow density $I_{inc}$, like the probability density $|\Psi_{tot}(x;k)|^2$, describes the whole
ensemble of particles. But now we come across another problem -- the substitution $I_{inc}$ for $I_{tot}$ is made here 'by hand' because $I_{inc}$
does not relate to $|\Psi_{tot}(x;k)|^2$ in the barrier region $[a,b]$.

Thus, unlike $v_{flow}^{free}$, the 'velocities' $v_{flow}^{(1)}$ and $v_{flow}^{(2)}$ have no clear physical meaning when $R(k)\neq 0$. None of
them can be unambiguously ascribed to the whole ensemble of particles or to its transmitted part. As a consequence, the physical meaning of the
dwell times $\tau_{dwell}^{(1)}$ and $\tau_{dwell}^{(2)}$ defined on the basis of these 'velocities' is unclear, too (see Sections \ref{larmor}
and \ref{Davi}).


\subsection{The asymptotic group time} \label{group}

The next important concept in the TTL is the Wigner time or group delay, often referred to as the phase time. By the nature, this time scale is an
asymptotic (extrapolated) group time. Its analysis is of great importance because, as is claimed, it has been experimentally verified (see Section
\ref{experiment}).

Following the Wigner time concept, we shall consider the case when $l_0\gg d$, assuming that at the initial time $t=0$ the wave packet peaks in
the $k$-space at the point $k$. As in the previous section, we begin with the free motion. In this case $\Psi_{tot}(x,t)$ coincides with
$\Psi_{inc}(x,t)$ (see Exp. (\ref{5})) extrapolated onto the whole $OX$-axis. The position $<\hat{x}>_{free}$ of the CM of this wave packet, for
any value of $t$, is given by the expression $<\hat{x}>_{free}=\hbar kt/m$. Thus, for the time $\tau_{group}^{free}$ spent by the CM in the region
$[a,b]$ we have $\tau_{group}^{free}=md/\hbar k$. That is, in this case the group time $\tau_{group}^{free}$ coincides with the dwell time
$\tau_{dwell}^{free}$ (see Exp. (\ref{7})); both give the transit time to describe the free wave dynamics in the region $[a,b]$.

The situation changes drastically when we proceed to the OCS. The first problem to appear in this case is that for $R(k)\neq 0$ the transmitted
part of the scattering wave packet is seen only at the final stage of scattering, far from the barrier, in the transmission region. Thus, in the
best case, the {\it group tunneling} time can be introduced within the standard model of the OCS only for some asymptotically large spatial
region, e.g., for the interval $[0,b+L]$ where $L\gg l_0$. That is, in the best case, it might be defined as an {\it asymptotic} group time.

However, even this cannot be done properly within the conventional model of the OCS (Section \ref{smodel}). Indeed, this approach allows us to
define correctly the time of arrival $\tau_{tr}^{ar}(x)$ of the CM of the transmitted wave packet at the point $x$ located far behind the barrier.
For this purpose we can use the explicit expression for the position $<\hat{x}>_{tr}(t)$ of the CM of the transmitted wave packet,
\begin{eqnarray*}
<\hat{x}>_{tr}(t)=\frac{\hbar kt}{m} -J^\prime(k)+d;
\end{eqnarray*}
here the prime denotes the derivative on $k$. "However, before the arrival time is related to a "transit time" one must know the departure time of
the thing that arrived" \cite{Win}. In the existing approaches it is assumed that the departure times of the CMs of the transmitted and reflected
wave packets coincide with each other (note that in the original Wigner's approach \cite{Wig} there is no necessity in this assumption, because
Wigner deals in \cite{Wig} with the problem of scattering a particle on a point-like scatterer, where there is only one scattering channel --
reflection). Since the CM of the incident wave packet starts from $x=0$, in our setting of the problem, the arrival time
$$\tau_{tr}^{ar}(b+L)=\frac{m}{\hbar k}\left(J^\prime(k)+L+a\right)$$ gives immediately the group transit time for the interval
$[0,b+L]$. Without the contributions of the outer regions $[0,a]$ and $[b,b+L]$, this expression yields the Wigner (asymptotic group) time
$\tau_W$ and corresponding delay time $\tau_{del}$:
\begin{eqnarray} \label{10}
\tau_W(k)=\frac{m}{\hbar k}J^\prime(k) \ppp \tau_{del}(k)=\tau_W-\tau_{group}^{free}=\frac{m}{\hbar k}\left[J^\prime(k)-d\right].
\end{eqnarray}

However, one has to remember that the above mentioned assumption about the departure time ignores the well known fact (see \cite{La2}) that there
is no causal relationship between the transmitted and incident wave packets. So that the Wigner time concept violates the causality principle.
And, it is unimportant in this case, whether Eq. (\ref{10}) leads to superluminal velocities or no.

\subsection{The "non-coherent flux-separation" timekeeping procedure} \label{flux}

As is seen, the main shortcoming of the preceding timekeeping procedure is its inability to distinguish the dynamics of transmitted and reflected
particles at the initial stage of scattering. In this connection, of importance is to dwell on "the non-coherent flux-separation" technique which,
as is claimed in \cite{Olk1}, resolves this problem. The fact is that, in reality, this approach does not solve this problem.

Firstly, the time operators $\hat{t}_{\pm}(x)$ introduced in this approach imply averaging over time, and the authors claim that such averaging
agrees with the conventional QM. However, in fact, this was proven only for the initial and final stages of the OCS, i.e., for the asymptotically
distant (from the barrier) spatial regions, when the total flux $J$ is equal either to $J_+$ or $J_-$; $J_\pm=J \Theta(\pm J)$ (see p.137); here
the positive $J_+$ and negative $J_-$ fluxes describe forward and backward motion, respectively.

Secondly, this formalism like the Wigner time concept violates the causality principle. Indeed, the probabilities $\rho_>(x_i,t)$ and
$\rho_>(x_f,t)$ defined for the remote points $x_i$ and $x_f$ to lie on the different sides of the barrier describe the one-particle ensembles
between whom there is no causal relationship. As a result, the time scale $\left\langle \tau_T(x_i,x_f)\right\rangle = \left\langle
t_+(x_f)\right\rangle -\left\langle t_+(x_i)\right\rangle$, defined as the "differences between the mean times referring to the passage of the
final and initial wavepackets through the relevant space-points", in fact bears no relation to dwelling the subensemble of {\it tunneling}
particles in the region $[x_i,x_f]$. In this approach the departure time $\left\langle t_+(x_i)\right\rangle$ does not describe the subensemble of
{\it to-be-transmitted} particles whose time of arrival at the point $x_f$ is described by $\left\langle t_+(x_f)\right\rangle$.

\subsection{Time scales associated with total and partial densities of states} \label{PDS}

In this section we consider the timekeeping procedure \cite{Gas,But2} to define the characteristic times of the OCS through the partial densities
of states (PDOSs). These quantities appear within the scattering-matrix formalism to describe the response of the system under study to the
infinitesimal variation of the potential $V(x)$. Again, this approach is of interest because, as is claimed in \cite{Gas}, PDOSs carry the
information not only about the future of scattering particles, but also about their past.

Let the OCS be characterized by a scattering matrix with elements $S_{\alpha\beta}$, where the indices $\alpha$ and $\beta$ label, respectively,
outgoing and incoming scattering channels of the system under study (see \cite{Gas}). The local PDOS $d\eta_{\alpha\beta}/dE$ are written in
\cite{Gas} in the form
\begin{eqnarray} \label{11}
\frac{d\eta_{\alpha\beta}}{dE}(x)\equiv -\frac{1}{4\pi i}\left(S^\dag_{\alpha\beta} \frac{\delta
S_{\alpha\beta}}{\delta V(x)}-\frac{\delta S^\dag_{\alpha\beta}}{\delta V(x)}S_{\alpha\beta} \right)
\end{eqnarray}
where the off-diagonal PDOSs are always positive; $\delta /\delta V(x)$ denotes a functional derivative.

Then the injectivity $d\eta^{inj}_\beta/dE$ of the incoming channel $\beta$ as well as the emissivity $d\eta^{emis}_\alpha/dE$ into the outgoing
channel $\alpha$ are
\begin{eqnarray} \label{12}
\frac{d\eta^{inj}_\beta}{dE}(x)=\sum_{\alpha}\frac{d\eta_{\alpha\beta}}{dE}(x),\ppp
\frac{d\eta^{emis}_\alpha}{dE}(x)=\sum_{\beta}\frac{d\eta_{\alpha\beta}}{dE}(x).
\end{eqnarray}
The PDOSs, injectivity and emissivity enter into the decomposition of the total DOS as follows
\begin{eqnarray} \label{13}
\frac{d\eta}{dE}(x)=\sum_{\alpha\beta}\frac{d\eta_{\alpha\beta}}{dE}(x)=\sum_{\beta}\frac{d\eta^{inj}_\beta}{dE}(x)=
\sum_{\alpha}\frac{d\eta^{emis}_\alpha}{dE}(x).
\end{eqnarray}
As was said in \cite{Gas} about PDOSs, "They are based on both a preselection and postselection of carriers, i.e., they group carriers according
to the asymptotic region from which they arrive ($\beta$) and according to the asymptotic region into which they are scattered ($\alpha$). We
emphasize that the PDOSs are mathematical constructions. Whether these quantities are by themselves of physical relevance might well depend on the
problem under investigation."

Note that in the case of the OCS, for scattering channels located at the left and right sides of the barrier region, we have $\alpha,\beta=1$ and
$\alpha,\beta=2$, respectively (see \cite{Gas}). Thus, when a particle impinges on the barrier from the left, the relevant local PDOSs are
$d\eta_{11}/dE$ and $d\eta_{21}/dE$, respectively.

As was shown in \cite{Gas}, the corresponding injectivity $d\eta^{inj}_1/dE$ determines the time scale $d\tau_1$:
\begin{eqnarray} \label{14}
d\tau_1(x)=\frac{|\Psi_{tot}(x)|^2}{I_{inc}}dx=2\pi\hbar \frac{d\eta^{inj}_1}{dE}(x)dx.
\end{eqnarray}
That is, $d\tau_1/dx$ coincides with the dwell time $\tau_{dwell}^{(2)}$ discussed in Section \ref{dwell}; both these quantities might be relevant
in the case of the ONCS.

In some cases the local PDOSs are connected to the local Larmor times (see expressions (57--60) in \cite{Gas}) which are considered in \cite{Gas}
as "physically well-defined quantities" to describe the OCS. According to the authors of the paper, "The results (57)--(60) connect the local PDOS
with physically well-defined quantities, which indicates the relevance of the PDOS". But this is not. The physical relevance of the PDOSs in
studying the temporal aspects of the OCS is moot.

Firstly, the PDOSs $d\eta_{21}/dE$ and $d\eta_{11}/dE$ to enter Exps. (57--60) connect the outgoing channels $\beta=2$ and $\beta=1$ with {\it the
same incoming channel} $\alpha=1$. This means that none of these outgoing channels, taken alone, is linked causally to this incoming channel.
Incoming channels that would be causally linked to either of these two outgoing channels are unknown within the conventional model of the OCS (it
is this problem that is under study in our approach \cite{Ch6,Ch1,Ch2,Ch3} (see Section \ref{separ})). So that the PDOSs $d\eta_{21}/dE$ and
$d\eta_{11}/dE$ have no relation to transmission or reflection.

Secondly, within the conventional model of the OCS, the Larmor-clock procedure is internally inconsistent. Revealing the main shortcomings of the
existing Larmor-clock procedure is our next goal.

\subsection{Larmor times} \label{larmor}

Initially proposed in the works \cite{Baz2}, this procedure was developed further by Buttiker \cite{But}. Its main idea is as follows. An
infinitesimal magnetic field directed along the $OZ$-axis is confined to the barrier region $[a,b]$ on the $OX$-axis. At the initial time $t=0$ a
beam of electrons scattering on the potential barrier is in the quantum state to represent the statistical mixture of two subensembles of
particles with the $z$-th spin components $\hbar/2$ and $-\hbar/2$. This state is assumed to be such that the electron spin averaged over the
mixture is strictly orthogonal at $t=0$ to the magnetic field and the direction of the motion of particles. Outside the barrier the spin is
constant. When electrons enter the barrier region the average spin starts a Larmor precession. When they leave the barrier the precession stops.

In this timekeeping procedure the average spin of particles plays the role of a clockwise. For transmitted particles the final position of the
clockwise coincides with the direction of the electron spin averaged over the transmitted portion of the scattered beam. Its initial position
coincides, as is assumed in \cite{But}, with the direction of the spin averaged over the {\it whole incident} beam, which, as is said above, is
strictly orthogonal to the magnetic field and the velocity of particles.

As is shown in \cite{But}, in the course of scattering, the average spin of transmitted particles not only rotates, due to the Larmor precession,
in the plane orthogonal to the magnetic field, but also acquires a nonzero $z$-th component. As a consequence, the Larmor timekeeping procedure
provides two independent characteristic times for transmission: $\tau_\bot^{tr}$ associated with the Larmor precession, and $\tau_\|^{tr}$
associated with the emergence of a spin component parallel to the magnetic field (see also Section \ref{f4}). That is, figuratively speaking, by
Buttiker there are in fact two Larmor clocks associated with transmission: one clock measures the Larmor time $\tau_\bot^{tr}$ that is precisely
the dwell time $\tau_{dwell}^{(2)}$; another measures the quantity $\tau_\|^{tr}$ that determines the so-called traversal time to coincide in the
opaque limit with the B\"{u}ttiker-Landauer time \cite{La2}.

Note that the effect of aligning the average spin with the magnetic field was found for reflected particles too. As was shown in \cite{But}, the
corresponding Larmor time $\tau_\|^{ref}$ is such that $T\tau_\|^{tr}+R\tau_\|^{ref}=0$ for any given energy $E$.

However, the equality $\tau_\bot^{tr}=\tau_{dwell}^{(2)}$ is paradoxical in essence. Indeed, it says that the dwell time $\tau_{dwell}^{(2)}$
defined in terms of the total wave packet to move in the barrier region (see Section \ref{dwell}) turns out to coincide with the time scale
$\tau_\bot^{tr}$ defined in terms of the transmitted wave packet to move far from the barrier region.

The emergence of the nonzero scattering times $\tau_\|^{tr}$ and $\tau_\|^{ref}$ is even more paradoxical. The point is that, according to the
assumption made in \cite{But}, the $z$-th component of the average spin for both subprocesses is zero at the initial time. Thus, this spin
component, as a motion integral in this scattering problem, must be zero at all stages of scattering. At the same time the Larmor procedure
violates this requirement. Moreover, this "effect" appears for the reflection subprocess even when the infinitesimal magnetic field is switched on
far from the barrier region, in the transmission zone. As was said in \cite{Leav,Leav1} in this connection, "the Larmor-clock approach leads to a
result contrary to the common sense notion that a reflected particle does not spend any time on the far side \ldots of the potential barrier". (As
is shown in \cite{Ch2} (see also Section \ref{f4}), the $z$ components of the average spins of transmitted and reflected particles are {\it
nonzero} at the initial time and the "interactions" times $\tau_\|^{tr}$ and $\tau_\|^{ref}$ are, in fact, the initial positions of the Larmor
clocks for transmission and reflection, respectively. These quantities remain constant in the course of scattering and, thus, do not measure the
duration of these subprocesses.)

These results cannot be considered as well-established. There are two steps in the Larmor-clock procedure, which undermine its legitimacy. The
first one is that "The polarization of the transmitted (and reflected) particles is compared with the polarization of the incident particles"
\cite{But}. But this step is evident to contradict the observation \cite{La2} that there is no causal relationship between the transmitted
(reflected) and incident particles. Thus, like the Wigner time this concept violates the causality principle.

Another unjustified assumption concerns the dynamics of the average spin of transmitted particles in the plane orthogonal to the magnetic field.
As is assumed in \cite{But}, in the barrier region the spin experiences only the (smooth) Larmor precession in this plane. But this assumption is
justified only for the spin averaged over the {\it whole} beam of particles, whose state experiences the unitary quantum evolution at all stages
of scattering. At the same time, transmission is only a part of the OCS; extension of properties of the OCS onto its subprocesses is unjustified
(see also Section \ref{time-dep}).

Thus, the implicit assumption made in \cite{But} about the unitarity of the tunneling dynamics in the barrier region is unjustified, and one
should not exclude that, apart from the Larmor precession, other physical effects could alter the average spin of transmitted particles in the
barrier region (see Section \ref{f4}). To clearly answer this question, one has to reveal the dynamics of transmitted particles at all stages of
scattering.

Note that all this fully concerns the recent versions \cite{Dav,Lun} of the Salecker-Wigner-Peres procedure \cite{Sal,Per}. As analogs of the
Larmor-clock procedure, they suffer from the same drawback: in all these timekeeping procedures quantum clocks are coupled, at the initial and
final stages of scattering, with ensembles of particles, which are not linked causally to each other.

\subsection{Davies' clock-based timekeeping procedure} \label{Davi}

Here we dwell in short on one important peculiarity of Davies' procedure  \cite{Dav}. The main idea of introducing the transmission time in
\cite{Dav} is as follows:
\begin{quote}
"To achieve this, the particle is coupled (weakly) to a quantum clock. The coupling is chosen to be non-zero only when the particle's position
lies within a given spatial interval\ldots Initially the clock pointer is set to zero. After a long time, when the particle has traversed the
spatial region of interest with high probability, the position of the clock pointer is measured. The change in position yields the expectation
value for the time of flight of the particle between the two fixed points."
\end{quote}

This clock-based procedure has been designed for timekeeping a {\it scattering} particle, but its key features have been illustrated in \cite{Dav}
by the example of a {\it free} particle. That is, per se, the dynamics of free and scattering particles are treated in \cite{Dav} as qualitatively
identical. But this is not; this timekeeping procedures (and all other ones in the TTL), being true in the case of a free particle, violates the
causality principle when it is used for a particle scattering by a potential barrier.

To see the principal difference between the timekeeping of the quantum dynamics of free and scattering particles, one has to take into account the
following two things. First, the mentioned coupling between a quantum clock and a particle in the (quasi)stationary state is realized in
\cite{Dav} as the coupling between the clock and the phase of the wave function that describes the particle's state. Second, when "the spatial
region of interest" is empty the incident wave, that traverses it, never splits into parts. Whilst, when this region includes the potential
barrier, the incident wave splits here into two waves -- transmitted and reflected.

The latter means that at the initial stage of scattering the clock pointer is coupled in this model to the phase of the {\it incident} wave (that
describes the whole ensemble of particles, without distinguishing its to-be-transmitted and to-be-reflected parts), while at the final stage it is
coupled to the phase of the {\it transmitted} wave (that describes only the transmitted part of the incident ensemble). That is, again, at the
initial and final stages the clock pointer is coupled to causally disconnected ensembles of particles.

Note that Davies, when dealing with initial stage of scattering, prefers to speak of 'setting to zero of the clock pointer', rather than of
'measuring the departure time' as in the Larmor procedure. But nothing has been changed, in essence. Setting to zero of the clocks pointer in
\cite{Dav} is based on the implicit assumption that the average departure times of the whole ensemble of particles and its to-be-transmitted part
(causally connected to the transmitted subensemble), coincide with each other. But, as was said above, this is not obvious for causally
disconnected subensembles. So that setting to zero of clocks used for measuring the tunneling time was performed in \cite{Dav} with violating the
causality principle.

\section{About the controversy around the Hartman effect} \label{hart}

So, all the tunneling time concepts violate the causality principle and, at first glance, due to this fact there is no need to discuss what
follows from them. But this is not. The fact that these approaches use the wave packet $\psi_{inc}(x,t)$ for the introduction of the asymptotic
tunneling time, while ignoring the well-known fact that there is no causal link between the incident and transmitted wave packets $\Psi_{tr}(x,t)$
does not mean that these concepts necessarily lead to incorrect results. To reject these concepts as defective, one has also to prove that their
CMs really start at different time from the point $x=0$. As a consequence, we have at present a contradictory situation in the TTL: on the one
hand, nobody has (dis)proved this fact and hence cannot reject the results that follow from these concepts; on the other hand, because of the
logical flaws of these concepts nobody can unambiguously interpret their results. And a long-standing controversy around the interpretation of the
Hartman effect is the most striking example.

Our next step is to analyse this controversy in detail. However, before doing so, we have to recall that there are two versions of this effect.
The 'ordinary' Hartman effect \cite{Har} was found by the example of the rectangular barrier, on the basis of the Wigner group time. Its essence
is that the Wigner time saturates in the so called 'opaque limit', i.e., when $d\to \infty$ for $E<V_0$. In the case of the generalized Hartman
effect \cite{Ol3,Rec5}, found by the example of two identical rectangular barriers (whose height and width equal to $V_0$ and $d$, respectively),
this characteristic time becomes independent, in the opaque limit, not only on the width of the barriers, but also on the distance between them.



In the physical community, the group velocity associated with the Wigner tunneling time is usually interpreted as the average velocity of
tunneling particles in the barrier region, and one of the central issues in the TTL has been to reconcile the superluminal group velocities
measured in the optical tunneling-time experiments (see Section \ref{experiment}) with SR. Now there is a number of ideas of solving this problem,
but none of them can be considered as commonly accepted. In this connection, our next task is to show why even the most prominent ideas of
reconciling the Hartman effect with SR, in reality, did not reach this aim.

\subsection{On the 'reshaping argument', signal velocity and Kramers-Kronig relations} \label{hart}

We begin with the most popular idea (see, e.g., \cite{Ste,Ste2,But3,Soc1,Chen,Wang,Japh}) according to which there is nothing paradoxical in the
Hartman effect, and the tunneling phenomenon does not contradict SR. This idea contains three main ingredients: the so called 'reshaping
argument', the 'signal velocity' argument and the 'dispersion relations' argument.

As was said in \cite{Ste2},
\begin{quotation}
(\i) "In classical optics, the existence of group velocities greater than $c$, and even negative ones under certain conditions, is known, and has
been observed experimentally. This phenomenon is understood as a "pulse reshaping" process, in which a medium preferentially attenuates the later
parts of an incident pulse, in such a way that the output peak appears shifted towards earlier times\ldots"
\end{quotation}
\begin{quotation}
(\i\i) "Although the apparent tunneling velocity $(1.7\pm 0.2)c$ is superluminal, this is not a genuine signal velocity, and Einstein causality is
not violated."
\end{quotation}

The mentioned tunneling time experiments will be analysed in Section \ref{experiment}, and here we focus our attention only on this interpretation
of the Hartman effect. Having been developed twenty years ago, this interpretation (or, at least, its ingredients such as the 'signal velocity'
and 'dispersion relations' arguments) remains popular to this day (see, e.g., \cite{Pol,Pol1,Lyp,Lett,Gus}). At the same time it suffers from
logical flaws. To show them, let us dwell on the above two quotes in detail.

{\bf (\i) "\ldots This phenomenon is understood as a "pulse reshaping" process, in which a medium preferentially attenuates the later parts of an
incident pulse, in such a way that the output peak appears shifted towards earlier times\ldots"}


However, both in QM and CED the Hartman effect is associated with an {\it elastic} scattering process. This implies that layered photonic
structures, where this effect is observed, consist only of {\it passive} (non-absorptive, non-active) media. Thus, the phrase "a medium
preferentially {\it attenuates}" is misleading in this case: the layer of a passive medium which plays the role of a photonic barrier splits the
incident light pulse into parts, rather than attenuates its transmitting part.

Recall that the main peculiarity of the Wigner group time is that, within the timekeeping procedure to underlie this concept (see Section
\ref{group}), the time of arrival at the point $x=b+L$ was defined for the {\it transmitted} wave packet, while the departure time from the point
$x=0$ was defined for the {\it incident} wave packet which is causally disconnected with the former (see \cite{La2}. In fact, the 'reshaping
mechanism' is intended to describe the process of 'transformation' of the incident pulse into the transmitted one. But there is no causal
relationship between these peaks. The incident peak transforms into the {\it sum} of the transmitted and reflected peaks, rather than only into
the transmitted peak. Thus, 'reshaping mechanism' violates the causality principle, a priori. In this case, the value of the tunneling velocity
(whether it is superluminal or subluminal) is of secondary importance.

The internally conflicting character of the logics to underlie "pulse-reshaping argument" is seen from the statement, which most briefly and
precisely expresses the essence of this argument: "\ldots the causality is not violated since reshaping destroys causal relationship between the
incident and the transmitted peaks" \cite{Soc1}.

{\bf (\i\i) Although the apparent tunneling velocity\ldots is superluminal, this is not a genuine signal velocity, and Einstein causality is not
violated:}

By a genuine signal velocity is meant here (see \cite{Ste2}) the propagation velocity of an abrupt leading edge of a light pulse tunneling through
the barrier (see also \cite{Pol,Pol1,Lyp,Lett,Gus}); the spectra of such pulses are always unbounded in the high-frequency limit. As is said in
\cite{Ste}, "Fronts are preserved in the output. Therefore, although there is no physical law which guarantees that an incoming peak turns into
outgoing peak, there is a physical law namely causality, that guarantees that an incoming front turns into an outgoing front, even when the front
carriers little energy or probability".

To show that the signal (or front) velocity is always subluminal, Chiao and Steinberg \cite{Ste} offer the idealized model of a "black box" which
locally links an input to an output wave form by means of a linear transfer function. They show that the relationship between the input and output
is causal in this model when the Fourier transform of this transfer function obeys the Kramers-Kronig relations. And they claim then, with
referring to Jackson \cite{Jack}, that "the generalization of this argument to {\it propagation} through any {\it spatially extended} "black box",
that is linear and causal, is straightforward".

However, the reference to Jackson \cite{Jack} is inappropriate here, because no part in this textbook concerns the problem under consideration. At
first glance, it is the exercise 7.8 on the page 234 that has a direct bearing on this problem. Indeed, this exercise is posed as follows:
\begin{quotation}
"A very long plane-wave train of frequency $\omega_0$ with a sharp front edge is incident normally from vacuum on a semi-infinite dielectric
described by an index of refraction $n(\omega)$ and occupying the half-space $x> 0$. Just outside the dielectric (at $x = 0$) the {\it incident}
electric field is $E_0(0,t)=\theta(t)e^{-\epsilon t}\sin(\omega_0 t)$, where $\theta(t)$ is the step function\ldots The exponential decay constant
$\epsilon$ is a positive infinitesimal\ldots;

(b) Prove that a sufficient condition for causality (that no signal propagate faster than the speed of light in vacuum) in this problem is that
the index of refraction as a function of {\it complex} $\omega$ be an analytic function, regular in the upper half $\omega$ plane with
nonvanishing imaginary part there, and approaching unity for $|\omega| \to\infty$.

(c) Generalize the argument of (b) to apply to any incident wave train."
\end{quotation}
However, as is seen, the boundary condition for $E_0(0,t)$ at the point $x=0$ does not correspond to the phrase "a long plane-wave train\ldots
with a sharp front edge is {\it incident normally from vacuum on a semi-infinite dielectric}\ldots" In reality, this exercise deals with the wave
field which is generated at the left boundary $x=0$ of the semi-infinite dielectrics and propagates into the region $x>0$ occupied by this
dielectric. Unlike the OCS, there is no reflection here and, thus, there is no splitting of the incident wave packet into two coherently evolved
parts. So that the problem considered in this exercise has nothing in common with that concerned in the statement (\i\i).

Note that Sokolovski \cite{Soc1} unlike Chiao and Steinberg refers, for supporting the statement (\i\i), to the chap. 3 of the book \cite{Baz}.
However, in our opinion this reference, too, is misleading. The 'dispersion-relation argument' is applied in \cite{Baz} only to {\it one-}channel
scattering processes, when an incoming pulse does not split within "a black box" into several outgoing channels. While, in the case of a
non-resonant tunneling, we deal with a two-channel scattering process, and hence the dispersion-relation argument is insufficient here for proving
or disproving the statement that the transmission channel is governed by the Einstein causality.

\subsection{"Tunneling confronts special relativity"} \label{Nimtz}

The privileged status of the signal velocity was put in doubt by Nimtz and Haibel (on some problems associated with the front velocity, see also
\cite{Brun}) who stressed in \cite{Nim1} that "A physical transmitter produces signals of finite spectra only\ldots [Hence f]ront of a signal has
no physical meaning\ldots Only the complete envelope\ldots is the appropriate signal description". Thus, according to Nimtz and Haibel, namely the
group velocity is a signal velocity. And Nimtz concluded that "Tunneling confronts special relativity" \cite{Nim}; tunneling takes place due to
"virtual particles" \cite{Nim2}.

This explanation of the Hartman effect, which actually comes out beyond the scope of the 'usual' special relativity, coincides in many respects
with the one presented in \cite{Olk1}, which is based on the co called 'non-restricted special relativity' (NSR) \cite{Rec6,RecF}. The authors of
this approach treat a superluminal tunneling as analogue of the superluminal motion of the so called 'X-waves' -- real solutions to the Helmholtz
equation \cite{RecF}. In this case, superluminal tunneling obeys the causality principle of the non-restricted SR, with its "switching rule" or
"reinterpretation principle" (see \cite{Rec6,RecF} as well as \cite{Mal}).

Both these approaches to the Hartman effect are internally consistent and at this point, when the tunneling dynamics is still unknown, we cannot
disprove them. However, one remark is of importance for our further study. It concerns the nature of X-waves. As is seen from \cite{RecF}, one has
to distinguish two cardinally different types of problems for any wave equation (equivalent to the Helmholtz one), where such solutions as X-waves
appear: problems where X-waves are generated by a single source as well as problems where they are generated by two coherently operating sources.
The first case happens, for example, "for a plane that moves in the air with constant supersonic speed" \cite{RecF}. The second one occurs when a
X-wave represents a superposition of two subsonic waves running from two coherent sources. Both these problems are realistic. However, in CED, the
realization of the first case requires either a hypothetical particle (tachyon) or virtual particles. Whether this case relates to the tunneling
dynamics is an open question, as long as this dynamics remains unknown at all stages of scattering.

\subsection{On Winful's reinterpretation of the Wigner tunneling time} \label{win}

One more way to resolve the conflict between the current description of the OCS and SR is to recognize the Hartman effect as an artifact of an
internally inconsistent tunneling time concept. This argument, put forward by the authors of \cite{Nus,Win}, is justified because the above
approaches leave unsolved the problem of distinguishing the transmission and reflection dynamics at the initial stage of scattering (the latter is
important for a proper determination of the departure time for particles which eventually are transmitted by the barrier). At the same time, we do
not agree both with the authors of \cite{Nus} who claim that the TTP is an ill-posed problem and with Winful who believes that the Wigner time
admits reinterpretation. Our arguments in favour of physical relevancy of the TTP are presented in Section~\ref{req}. Here we dwell on Winful's
idea.

Winful stresses in \cite{Win} that "Wave propagation in any medium (including vacuum) proceeds through the storage and release of energy". That
is, he prefers to consider the transfer of the electromagnetic energy through a photonic barrier as its accumulation in the barrier region and the
subsequent outflow from this region. He divides the process of transferring the electromagnetic energy through the barrier region into two stages
-- the energy accumulation and its subsequent release. This implies that the duration of the energy transfer can now be defined as the sum of the
"accumulation time" and the "release time".

At the same time Winful associates the duration of the energy transfer only with "\ldots a lifetime of stored energy {\it leaking out} of both
ends of the barrier\ldots". He specifies that its duration "\ldots is not the time it takes for the input peak to propagate to the exit since the
pulse does not really propagate through the barrier\ldots What is really measured is the lifetime of stored energy {\it escaping} through both
ends" (italics supplied). That is, in fact, the accumulation stage turns out to be beyond the framework of Winful's timekeeping procedure -- the
process of transferring the electromagnetic energy through the barrier region is changed in this procedure by that of escaping the stored energy
from this region.

Thus, Winful's reinterpretation of the Wigner (transit) time is moot when one deals with the OCS. But it might be useful in the case of the ONCS.
(see Section \ref{smodel}).

\section{On the Bohmian approach to the tunneling time problem} \label{Bohm}

The Bohmian approach to the TTP requires a particular attention, because the Bohmian model of the OCS is considered as a model that sees the
transmission and reflection dynamics at all stages of this process (see, e.g., \cite{Leav,Kre}): Bohmian trajectories defined for transmitted and
reflected particles occupy non-overlapping spatial regions at all stages of scattering.

Of course, if this model were valid, there would be no need for this article. But this is not. The ability of the modern Bohmian model to see the
whole dynamics of the subprocesses is delusive. To show this, let us dwell on the main points of this model. The wave function $\Psi_{tot}(x,t)$
is presented here (see, e.g., \cite{Hol}) in the form $\Psi_{tot}(x,t)=M_{tot}\exp(iS_{tot})$ and the Schr\"odinger equation transforms into two
real equations
\begin{eqnarray} \label{1b}
\frac{\partial S_{tot}}{\partial t}+\frac{1}{2m}(\nabla S_{tot})^2-\frac{\hbar^2}{2m}\frac{\nabla^2 M_{tot}}{M_{tot}}+V=0,\ooo \frac{\partial
M_{tot}^2}{\partial t}+\nabla\left(\frac{M_{tot}^2\nabla S_{tot}}{m}\right)=0;
\end{eqnarray}
here $M_{tot}^2$ is the probability density and $j_{tot}(x,t)=M_{tot}^2 \nabla S_{tot}(x,t)/m$ is the probability current density. According to
this approach, $\nabla S_{tot}(x,t)/m$ is the trajectory velocity.

The main peculiarity of this model is that Bohmian trajectories, as the flow lines of $j_{tot}(x,t)$, do not intersect each other. As a
consequence, two sets of trajectories ending in the non-overlapping transmission and reflection zones are localized in macroscopically distinct
spatial regions at all stages of scattering, including the initial stage.

At first glance, owing to this property, the Bohmian model provides a solid basis for solving the TTP. However, this is not. Yes, this model (see,
e.g., \cite{Kre,Leav,Leav1}) does not lead to the Hartman effect. But it leads to other paradoxes (see \cite{Leav1}). And what is more important,
there is a reason to consider the separation of transmission and reflection in this model as incorrect. The point is that the probability current
density $j_{tot}$ and probability density $M_{tot}^2$ in Eqs. (\ref{1b}) possess by mutually exclusive properties: while the former allows one to
see the transmission and reflection dynamics at all stages of scattering, the latter does not (as was stressed in \cite{Nus}, "transmission and
reflection are inextricably intertwined"). Thus, the modern Bohmian model of the OCS cannot be considered as well-established.

Of course, this fact does not at all mean that the Bohmian approach is invalid by itself. To some extent, the well-known wave-packet analysis and
Bohmian approach (if one considers the Bohmian trajectories as merely the flow lines, and nothing more) are two sides of the same 'coin' -- QM.
The former visualizes the results of monitoring the quantum probability density associated with a time-dependent wave function $\Psi_{tot}(x,t)$,
and the latter visualizes those of monitoring the corresponding probability flow density. Thus, the shortcomings of the Bohmian model of the OCS
result, in fact, from the pathological properties of the wave function $\Psi_{tot}(x,t)$ that represents a micro-cat state at the final stage of
scattering. So that the modern Bohmian model and the wave-packet analysis (see Section \ref{smodel}, both give internally inconsistent
descriptions of the OCS.

The Bohmiam approach, like a litmus test, helps us to reveal the weaknesses in the existing practice of quantum-mechanical description of
one-particle scattering states which represent micro-cat states. The example with the OCS shows that this practice is erroneous. In particular,
neither tunneling times nor one-particle trajectories can be correctly introduced for the whole ensemble of particles described by the wave
function $\Psi_{tot}(x,t)$.

\section{About 'weak' and 'direct' measurements of the tunneling time} \label{measure}

So, all the existing theoretical approaches to the TTP give no clear answer to its key question because they violate the causality principle (see
Section \ref{req}). A similar situation reigns in studying this problem on the experimental level: the current view on the role of 'tunneling
time' experiments in solving the TTP is contradictory, and the timekeeping procedures that underlie the tunneling time experiments \cite{Ste,Pol},
in which the Wigner group time was measured, suffer like the tested theoretical concept from the same shortcomings. To show this is our next goal.

Let us begin with the widespread view on the relationship between theoretical 'tunneling time' concepts and experiments designed for their
testing. According to this view experiment plays a crucial role in studying the temporal aspects of tunneling: the experimental confirmation of
any theoretical tunneling-time concept must be considered as a sufficient reason to treat it as a well-established concept. This is exemplified by
the following statements:
\begin{quote}
"The various candidates for general answers to this question [of the TTP] have also been critically examined. All have been found to suffer one
logical flaw or another, flaws sufficiently serious that must be rejected. [O]ne could turn to tunneling experiments now in progress with aim to
thoroughly understanding the temporal aspects of the individual experiments." \cite{Ha2}
\end{quote}
\begin{quote}
"There is no copyright on the expressions traversal times and tunneling time; each author can choose an interpretation. If an investigator wants
to associate it with the time required to write the Barden tunneling Hamiltonian on the blackboard, we cannot say that is wrong. We can only ask
if this is a fruitful view, and we can ask if it is relatable to experiment" \cite{La2}.
\end{quote}
\begin{quote}
"\ldots Many experiments, mainly in optics, have now been performed to measure the tunneling time, and the purely theoretical debate has been
transformed into one in which actual data can be brought to bear on the question. In the process, it has become clear that one must make a careful
{\it operational} definition of exactly {\it how} the measurement of the tunneling time is actually performed" \cite{Ste}.
\end{quote}

At the same time, the last statement in \cite{Ste} made on the page 347 contradicts to another one in \cite{Ste}, which is made on the page 400:
\begin{quote}
"when the [tunneling time] problem is studied\ldots we come up against one of the central problems of quantum mechanics -- the extent to which one
can discuss quantities which have not been measured directly, such as past history of a particle we observe at the present time\ldots In the case
of tunneling, there is no clear way to separate "to-be-transmitted" and "to-be-reflected" portions [of the ensemble of incident particles]\ldots"
\end{quote}

The key point in this statement is that it questions the possibility of a {\it direct} measurement of the tunneling time. In this connection, it
is important to consider how this problem is treated in the optical 'tunneling time' experiments as well as in the well-known 'weak measurement'
procedure.

\subsection{Tunneling time and "weak measurements"} \label{weak}

There is a widespread opinion (see, e.g., \cite{Ste,Soc,Ste3,So1,Aha1,Aha}) that the appearance of superluminal values of the group tunneling
velocity can be explained on the basis of the concept of 'weak measurements' giving 'weak values' of physical observables \cite{AhVa,AhVa1}. As is
said in \cite{Ste2}: "\ldots when a "weak measurement" \ldots is made on a subensemble defined both by state preparation and by a postselection of
low probability, mean values can be obtained which would be strictly forbidden for any complete ensemble".

According to the concept of 'weak measurement', "\ldots a quantum system between two measurements [is described] by two state vectors: the usual
one, evolving from the time of the latest complete measurement in the past, and the other one evolving backward in time from the time of the
earliest complete measurement in the future" p. 2315 in \cite{AhVa}. For some variable $A$ "such "weak measurement" of $A$ performed on an
ensemble of systems, which were preselected in a state $|\Psi_1>$ and were postselected in a state $<\Psi_2|$ will yield an outcome which we call
a weak value of $A$" \cite{AhVa1}:
\begin{equation} \label{140}
A_w=\frac{<\Psi_2|A|\Psi_1>}{<\Psi_2|\Psi_1>}.
\end{equation}
In this case, it is assumed that "\ldots for the intermediate time interval [between two measurements, preslection and postselection] we have a
complete symmetry under time reversal" (see p.12 \cite{AhVa1}).

It is widely recognized that the 'two-state' averaging procedure (\ref{140}) is alien to quantum mechanics (see, e.g. \cite{Ste4}). In particular,
being applied to the TTP, it leads to complex probabilities and tunneling times. About other problems associated with the concept of 'weak
measurement' see \cite{Sv1,Sv2}. Our aim is to show that 'weak measurement' does not really allow one to measure the tunneling time.

Let us try to apply this approach to the OCS, namely, to the subensemble of transmitted particles. At first glance, since this approach is based
on the idea of conditional probabilities, we could expect that in this case the preselected state $|\Psi_1>$ and postselected state $<\Psi_2|$
should be $|\Psi_{inc}>$ and $<\Psi_{tr}|$, respectively. But this contradicts the above requirements, according to which the states $|\Psi_1>$
and $<\Psi_2|$ describe the forward and backward evolutions of the same (postselected and simultaneously preselected) subensemble of particles,
which possesses "a complete symmetry under time reversal".

At the same time none of the proponents of the 'weak measurement' idea has proven that the time evolution of the postselected subensemble must
possess this symmetry like that of the whole quantum ensemble. In particular, in the case of the OCS this is wrong a priori: for any
semitransparent potential barrier there is no quantum dynamics which would be reversible in time and simultaneously described by one incoming wave
and one outgoing wave, as it should be for the transmission dynamics (see Section \ref{separ}). In this case the state $|\Psi_1>$ linked to the
postselected state $<\Psi_{tr}|$ "\ldots by a parity flip combined with a time reversal\ldots", as it was done in \cite{Ste4}, has no relation to
to-be-transmitted particles. We have to conclude that the mean value (\ref{140}) comprises probability distributions associated with mutually
incompatible statistical data. Thus, this averaging procedure has no physical meaning.

\subsection{Tunneling time and tunneling time experiments with single photons} \label{experiment}

Let us now proceed to the analysis of experiments designed to measure the tunneling time of quantum particles. Since the critical analysis of most
known tunneling time experiments has already been done in \cite{Win} we shall consider only the single-photon optical experiments \cite{Ste2,Pol}
(see also \cite{Ste5}) which are closest to the quantum-mechanical model considered in our paper (see Section \ref{smodel}).

The main intrigue associated with these experiments consists in the fact that they, as was claimed, allow a {\it direct} measurement of the
tunneling time. As was stressed in \cite{Ste5}, concerning the single-photon experiment \cite{Ste2}: "We presented the first direct time
measurement confirming that the time delay in tunnelling can be superluminal, studying single photons traversing a dielectric mirror". Thus,
bearing in mind the above cited reasonings on the page 400 of \cite{Ste} (see the nearest quotation prior to Section \ref{weak}), one could expect
that the problem of separating "to-be-transmitted" and "to-be-reflected" portions, in the case of tunneling, was solved in this experiment. But
this not the case.

The scheme of measuring the group delay in this experiment is as follows (see p. 708 in \cite{Ste5}): "It employs a two-photon source in which
pairs of photons are emitted essentially simultaneously. The advantage of using these "conjugate" particles is that after one particle traverses a
tunnel barrier its time of arrival can be compared with that of its twin (which encounters no barrier), thus offering a clear operational
definition of the tunneling time".

But all just the opposite, the usage of such "conjugate" particles is a serious disadvantage of this scheme, because the {\it ensemble} of these
particles is not the twin of the {\it subensemble} of particles traversing the barrier (the former is the twin of transmitted and reflected
particles taken together). Thus, the experimental data obtained for these two ensembles (see "Coincidence profiles with and without the tunnel
barrier" in fig.3 \cite{Ste5}) are incompatible with each other. Therefore, to deduce the group delay by comparing such data contradicts the
probability theory and this fully concerns the experiment \cite{Pol}: the "single-photon" tunneling-time experiments \cite{Ste2,Pol} create an
illusion of a {\it direct measurement} of the group delay for transmitted photons.

We have to stress that the "coincidence profiles" in this figure describe single-photon {\it ensembles}, rather than {\it single photons}. In fact
this experiment follows exactly the timekeeping procedure to underlie the concept of the Wigner time, and it is not surprising that experimental
data \cite{Ste2,Pol} "confirm" this internally inconsistent concept (see Section \ref{group} and \cite{Win,Wang}). All this fully concerns not
only the 'single-photon' experiments \cite{Ste2,Pol}, but also the optical experiments dealing with light pulses (see, e.g.,
\cite{Brun,Chen,Nim,Pol1,Lyp,Lett,Gus,Runf}): the 'beam' group tunneling time measured in these experiments coincides (provided that physical
conditions are the same) with the 'single-photon' group tunneling time measured in  \cite{Ste2,Pol}. One way or another, all they measure the
difference in the group delay for a wave packet weakly transmitted through the barrier and the reference wave packet traversing the same distance
in the absence of the barrier.

Of course, the "schedules of motion" of these different packets may accidentally coincide with each other at the initial stage of scattering. If
this would so, the presented in \cite{Ste2,Pol,Ste5} interpretation of this experiment were valid. But nobody proved that the time evolutions of
these two packets coincide with each other at this stage. Moreover, in order to prove this one has to know the "schedule of motion" of the
transmitted packet at the initial stage of scattering. This is evident to be impossible when the whole prehistory of the transmitted packet
remains unknown. Thus, strictly speaking, the tunneling time has been neither defined nor measured.

\section{The TTP as a problem that stands alongside with such fundamental problem of QM as the Schr\"odinger's-cat and EPR-Bohm paradoxes} \label{req}

As was said in \cite{Mug}, "A very important aspect, not technical but fundamental, is that the existing solutions [of the TTP], or even the
identification of the difficulties, are closely linked to particular interpretations of quantum mechanics\ldots [At the same time] no simple,
unambiguous, and quick resolution of all deep questions involved may be expected, since these concern our understanding of the emergence of the
classical world of "events" from the quantum world of possibilities. While many explanations have been proposed, we are far from a universal
consensus on how this emergence occurs."

The main peculiarity of the OCS is that the final state $\Psi_{tr}+\Psi_{ref}$, a coherent superposition of two states to occupy macroscopically
distinct spatial regions, represents a 'micro-cat' state. That is, this state is precisely of the same type as the states of a radioactive atom in
the Schr\"odinger's thought-experiment, an electron EPR-pair in the EPR-Bohm experiment and a particle in the two-slit experiment. Thus, the
difficulties appearing in solving the TTP are closely linked to the problem of reconciling the quantum-mechanical superposition principle with the
principles of a macroscopic realism \cite{Leg}. This problem is common for all quantum scattering phenomena where a micro-system is in a micro-cat
state.

Schr\"odinger, by resorting to an allegory, showed that, within the existing theory of micro-cat states, the superposition principle is
incompatible with the basic principles of classical physics. And, as is widely believed, the main lesson of Schr\"odinger's experiment is that the
superposition principle associated with micro-cat states must be reconciled with the principles of a macro-world at the level of {\it
macro-objects} whose mass is much larger than that of atoms. Putting it differently, this problem is understood by most physicists as the {\it
macro-objectification} (measurement) problem (see, e.g., \cite{Ghi,Sch,Joo}).

Note, while Schr\"odinger attempted to extend the quantum-mechanical superposition principle onto the macro-world, Bell, instead, attempted to
extend the fundamental law of classical physics -- the existence of LHVs -- onto the micro-world. He developed the probabilistic classical model
of the dynamics of the electron EPR-pairs to be in a micro-cat state and derived his famous inequality for the probability distributions to
describe the EPR-Bohm thought-experiments with differently oriented polarization beam splitters. In doing so, he assumed that there exists a set
of LHVs which is common for EPR-pairs in these experiments. Later it was shown that this inequality is violated in experiments and in QM. These
facts have been interpreted as a proof of the non-existence of LHVs in the nature, as well as a proof of the fact that the quantum logics is
incompatible with the classical one.

However, as it follows from other probabilistic approaches to Bell's inequality (see, e.g., \cite{Acc1,Acc,Khr1,Khr2,Phil,Hess}), there is a less
'fatal' reason which has no relation to the (non)existence of LHVs but leads too to the violation of this inequality. The point is that Bell-type
inequalities have been known in classical probability theory since the time of G.~Boole, where their violation means simply that these
inequalities comprise mutually incompatible statistical data (i.e., they do not belong to the same Kolmogorov probability space).

This directly concerns Bell's inequality derived for the probability distributions associated with differently oriented polarization beam
splitters used for detection of electrons. These experimental contexts are incompatible with each other and hence, from the viewpoint of classical
probability theory, there is nothing surprising in the fact that Bell's inequality is violated in experiments. Bell not simply assumes the
existence of LHVs for these incompatible experimental contexts. He assumes that these contexts are described by the {\it common} set of LHVs.
Thus, in the last analysis, the fact of violation of Bell's inequality in experiments means simply that Bell's probabilistic theory of the
EPR-Bohm experiment, based on {\it 'non-Kolmogorovian' LHVs}, contradicts classical probability theory. {\it No physical results can be inferred
from this incorrect probabilistic theory.}

However, Bell's theory survives despite this justified criticism. Why? This takes place because the nonexistence of LHVs is supported not only by
Bell's theory, but also by the modern quantum-mechanical models of quantum phenomena in which the states of micro-objects represent micro-cat
states. This concerns a radioactive atom in the Schr\"odinger-cat paradox, a particle in the two-slit experiment as well as a particle taking part
in the OCS. In each case, the squared modulus of the corresponding wave function describes mutually incompatible statistical data. However, in the
conventional models of these phenomena, this quantity is treated (contrary to classical probability theory) as the probability density. As a
consequence, these quantum models, like Bell's theory, make QM incompatible with classical physical theories.

From the viewpoint of classical theories these quantum models do not allow one, in principle, to unambiguously interpret the experiments
associated with these phenomena. And the whole history of QM confirms this: the two-slit experiment -- "the only mystery of QM", according to
Feynman -- remains unexplained; the Schr\"odinger-cat paradox, treated as a 'macro-objectification' problem, remains unresolved (all the existing
solutions of this problem conflict with each other and suffer from serious logical flaws (see, e.g., \cite{Leg})). The TTP associated with the OCS
stands alongside with these two mysterious problem of QM. Despite the long-term studies, this problem like the previous two remains unsolved up to
now.

Of great importance is that all these three problems are different modifications of the same quantum-mechanical problem of QM -- the problem of
the adequate interpretation and description of micro-cat states. To paraphrase Feynman, it is this problem that is "the only mystery of quantum
mechanics." Thus, in order to reconcile QM with classical physics one has not only to reinterpret Bell's inequality, but also to reconsider the
existing quantum-mechanical theory of micro-cat states.

The proponents of the existing theory say usually that it suffers from logical flaws only from the point of view of classical logics. But
classical logics they say does not apply to micro-cat states, and the only way to understand the corresponding quantum phenomena is to appeal to
Experiment -- the two-slit experiment must be explained on the basis of the complementarity principle, the cat paradox must be solved as the
measurement problem, and the definitions of the tunneling time must be 'operational'.

However, what we see by the example of the TTP. The main peculiarity of the transmission and reflection times is that these two physical
quantities can be measured only {\it indirectly}. The duration of each subprocess can be extracted only from scattering data obtained for both
scattering channels with the help of two macroscopic detectors situated on different sides of the barrier at large distance from it. Thus, to
gather these experimental data is only a part of any tunneling time experiment. One has also to unambiguously interpret them. What is evident can
{\it not} be done without adequate theoretical model of the OCS, which would allow tracing the transmission and reflection dynamics in the spatial
regions where they are inaccessible for a direct experimental observation!

Bohr suggested that "[in QM] the unambiguous interpretation of any measurement must be essentially framed in terms of the classical physical
theories" \cite{Bohr}. This means that the adequate model of the OCS and tested 'operational' tunneling-time concept must respect "the classical
physical theories" at least at the asymptotically large spatial distances from the barrier.

Firstly, it must respect classical probability theory. Thus, any averaging over experimental data obtained for transmitted and reflected particles
with help of the above two detectors, as being mutually incompatible from the point of view of probability theory, must be forbidden in QM too. In
other words, at the final stage of the OCS, when the total wave function $\Psi_{tot}(x,t)$ represents a micro-cat state built of the
macroscopically distinct states $\Psi_{tr}(x,t)$ and $\Psi_{ref}(x,t)$, only $|\Psi_{tr}(x,t)|^2$ and $|\Psi_{ref}(x,t)|^2$ can be interpreted as
the probability current densities.

Secondly, any timekeeping procedure must respect the causality principle. The minimal requirement of the latter is that timekeeping any
deterministic physical process in the spatial region of interest is possible, if only the dynamics of this process is known everywhere in this
region. This implies reconstructing the whole prehistory of the subensembles of transmitted and reflected particles by their final states
$\Psi_{tr}(x,t)$ and $\Psi_{ref}(x,t)$.

The current quantum-mechanical model of the OCS is obvious to violate both these requirements. It treats this process as a process, indecomposable
onto alternative subprocesses. The proponents of the standard approach claim that "From physical point of view it is obvious that only one wave
function, namely the solution of the scattering problem obeying the proper physical boundary condition ($\Psi_{tot}$ in our paper), should be used
for calculating any physical quantities \ldots No any part of the total wave function can be used for that."

At the same time the incompatibility of the model, based only on $\Psi_{tot}$, with classical probability theory is reflected in its internal
inconsistency: the functions $M_{tot}(x,t)$ and $j_{tot}(x,t)$ lead to contradictory inferences with respect to separating the transmission and
reflection dynamics at all stages of scattering (see Section \ref{Bohm}); they also do not allow one to unambiguously introduce the dwell time for
the whole ensemble of particles. As a consequence, this model provides no basis for the unambiguous interpretation of the tunneling time
experiments. The appearance of interpretation problems associated with the observed Hartman effect illustrates this. (Note that, the current model
of the ONCS is quite valid: now $\Psi_{tr}(x,t)+\Psi_{ref}(x,t)$ does not represent a micro-cat state, because $\Psi_{tr}(x,t)$ and
$\Psi_{ref}(x,t)$ overlap each other. Thus, now namely $|\Psi_{tr}(x,t)+\Psi_{ref}(x,t)|^2$ is the probability density.)

Thus, in order to reconcile the quantum-mechanical description of the OCS with the classical physical theories, we have to represent the OCS as a
combined process consisting of two inseparable from each other subprocesses which behave as mutually exclusive at least far from the barrier
region. The average values of physical quantities can be defined only for the subprocesses.

At the mathematical level, we have to represent the total time-dependent wave function $\Psi_{tot}(x,t)$ as a micro-cat state, i.e., as a
superposition of two 'subprocess wave functions' (SWFs) whose norms give in sum the norm of $\Psi_{tot}(x,t)$. And then, on the basis of these
SWFs, we have to define characteristic times for each subprocess. This formalism must provide the basis for an indirect measurement of the
transmission and refection times. Our next aim is to present such a model (see also \cite{Ch6,Ch1,Ch2,Ch3}).

\section{The OCS as a combination of two inseparable coherent alternative subprocesses} \label{separ}
\subsection{The stationary case} \label{st}

The alternative approach \cite{Ch6,Ch1,Ch2,Ch3} represents the OCS, for any semitransparent potential barrier $V(x)$ and function
$\mathcal{A}(k)$, as a complex quantum process consisting of two inseparable but indirectly distinguishable subprocesses -- transmission and
reflection. For this purpose the total stationary wave function $\Psi_{tot}(x;k)$ (\ref{1}) is decomposed, for any value of $k$, into a coherent
superposition of two SWFs $\psi_{tr}(x;k)$ and $\psi_{ref}(x;k)$ that describe, respectively, the transmission and reflection subprocesses in all
spatial regions. The uniqueness of this decomposition is provided by physically motivated requirements imposed on the SWFs.

The first requirement follows from the fact that at the final stage of the OCS the total wave function $\Psi_{tot}(x,t)$ represents (see
(\ref{5})) the superposition of two non-overlapping wave packets $\Psi_{tr}(x,t)$ and $\Psi_{ref}(x,t)$. Thus, to provide the fulfilment of this
property for any function $\mathcal{A}(k)$, the stationary SWFs must be such that, for any values of $x$ and $k$,
\begin{itemize}
\item[(a)] $\Psi_{tot}(x,k)=\psi_{tr}(x,k)+\psi_{ref}(x,k)$.
\end{itemize}
In this case, the time-dependent SWFs
\begin{eqnarray} \label{1f1}
\psi_{tr,ref}(x,t)=\frac{1}{\sqrt{2\pi}} \int_{-\infty}^\infty \mathcal{A}(k) \psi_{tr,ref}(x,k) e^{-iE(k)t/\hbar}dk
\end{eqnarray}
are evident to obey, for any value of $t$, the equality
\begin{eqnarray}\label{2f1}
\Psi_{tot}(x,t)=\psi_{tr}(x,t)+\psi_{ref}(x,t).
\end{eqnarray}

Next requirement is dictated by the very nature of the SWFs:
\begin{itemize}
\item[(b)] each SWF must have only one incoming wave and only one outgoing wave; the outgoing wave of $\psi_{tr}(x,k)$ is $a_{out}
e^{ik (x-d)}$ and that of $\psi_{ref}(x,k)$ is $b_{out} e^{ik(2a-x)}$ (see (\ref{1})).
\end{itemize}
As is known, for any semitransparent potential barrier $V_0$ there is no solution to the stationary Schr\"{o}dinger equation, which would be
everywhere continuous together with its first spatial derivative and simultaneously possess one incoming wave and one outgoing wave. Thus, we must
weaken standard requirements imposed on the continuity of SWFs in order to ensure, on the one hand, the existence of a nontrivial resolution of
the decomposition problem and, on the other hand, to provide a causal relationship between the incoming and outgoing waves in each SWF.

Namely, we believe that an incoming wave and outgoing wave of each stationary SWF represent two different solutions to the Schr\"{o}dinger
equation (for the same potential function $V(x)$), that are linked to each other at some spatial point $x_c(k)$, according to the following
requirement:
\begin{itemize}
\item[(c)] at the point $x_c(k)$ each SWF must be continuous together with the corresponding probability current density.
\end{itemize}

In line with the (b) and (c) requirements, the SWF $\psi_{ref}(x,k)$ must be zero in the region $x>x_c(k)$ because its incoming and outgoing waves
move to the left of the point $x_c(k)$. Thus, $\psi_{ref}(x,k)$ is a currentless wave function and the point $x_c(k)$, that plays in this model
the role of the extreme right turning point for reflected particles, coincides with some zero of this function. This zero must be causally
connected to the potential barrier and, thus, must be nearest to the barrier, among all zeroth of $\psi_{ref}(x,k)$. It should obey the following
requirement:
\begin{itemize}
\item[(d)] for any value of $k$ the point $x_c(k)$ must coincide with such a zero of the currentless wave function $\psi_{ref}(x,k)$,
at which the quantity $|dx_c(k)/dk|$ takes the least value on the set of zeros of this function.
\end{itemize}
The difference between the $k$ dependence of the sought-for zero and others is most noticeable in the limiting case $k\to\ 0$.

As was shown in \cite{Ch1,Ch3,Ch4}, for symmetric barriers this point coincides for any value of $k$ with the midpoint of the barrier region:
$x_c=(a+b)/2$; that is, $dx_c(k)/dk=0$ for such barriers. The SWFs for reflection and transmission, that obey the above requirements, read as
follows. For $x\leq a$
\begin{eqnarray*}
\psi_{ref}(x;k)=\Api_{ref}e^{ikx}+b_{out}e^{ik(2a-x)}, \ppp \psi_{tr}(x;k)=\Api_{tr}e^{ikx};
\end{eqnarray*}
for $a\le x\le x_c$
\begin{eqnarray*}
\psi_{ref}(x;k)=\left(PA^{in}_{ref}+
P^*b_{out}\right)f(x-x_c;k)\kappa^{-1}e^{ika}\nonumber\\
\psi_{tr}(x;k)=\left[A^{in}_{tr}P f(x-x_c;k)+ a_{out}Q^*g(x-x_c;k)\right]\kappa^{-1}e^{ika};
\end{eqnarray*}
for $x\ge x_c$
\begin{eqnarray} \label{17}
\psi_{ref}(x;k)\equiv 0,\ppp \psi_{tr}(x;k)\equiv \Psi_{tot}(x;k);
\end{eqnarray}
\begin{eqnarray} \label{18}
\Api_{ref}=b_{out}\left(b_{out}^*-a_{out}^*\right)= \sqrt{R}(\sqrt{R}+ i\sigma\sqrt{T})\equiv \sqrt{R}\exp\left(i\sigma\lambda\right),\\
\Api_{tr}= a_{out}\left(a_{out}^*-b_{out}^*\right)\equiv \sqrt{T}\exp\left[i\sigma\left(\lambda-\frac{\pi}{2}\right)\right];\nonumber
\end{eqnarray}
where $\lambda=\arctan\sqrt{T(k)/R(k)}$; $\sigma(k)=+1$, if $F(k)=0$ (see (\ref{3}) and (\ref{3a})); otherwise, $\sigma(k)=-1$.

Simple analysis shows that $a_{out}\left(a_{out}^*-b_{out}^*\right)=a_{out}^*\left(a_{out}+b_{out}\right)$ and
\begin{eqnarray*}
\Api_{tr}+\Api_{ref}=1,\ppp |\Api_{tr}|^2+|\Api_{ref}|^2=1.
\end{eqnarray*}
Besides, it is easy to show that $|\psi_{tr}(x-x_c;k)|$ is an even function for symmetric potential barriers. In the region $x_c<x<b$
\begin{eqnarray*}
|\psi_{tr}(x-x_c;k)|^2=\frac{T}{\kappa^2}\left[|P|^2 f^2(x-x_c;k)+|Q|^2 g^2(x-x_c;k)-(PQ^*+P^*Q)f(x-x_c;k)g(x-x_c;k) \right].
\end{eqnarray*}

And also, letting $\psi_{tr}(x;k)=M_{tr}(x;k)\exp[iS_{tr}(x;k)]$, we obtain that at the point $x_c$
\begin{eqnarray} \label{180}
S_{tr}(x_c-0;k)=S_{tr}(x_c+0;k), \ppp \frac{\partial S_{tr}(x;k)}{\partial x}\Bigg|_{x=x_c-0}=\frac{\partial S_{tr}(x;k)}{\partial
x}\Bigg|_{x=x_c+0},\\
M_{tr}(x_c-0;k)=M_{tr}(x_c+0;k), \ppp \frac{\partial M_{tr}(x;k)}{\partial x}\Bigg|_{x=x_c-0}=-\frac{\partial M_{tr}(x;k)}{\partial
x}\Bigg|_{x=x_c+0}\neq 0.\nonumber
\end{eqnarray}
That is, for symmetric potential barriers the values of the SWF $\psi_{tr}(x;k)$ and its first derivative $\partial\psi_{tr}(x;k)/\partial x$ in
the limit $x\to x_c-0$ can be calculated through those of the total wave function $\Psi_{tot}(x;k)$ and its first derivative at the point $x_c$.

\subsection{The time-dependent case}\label{time-dep}

The stationary wave functions $\psi_{tr}(x;k)$ and $\psi_{ref}(x;k)$ found for any value of $k$ lead to the unique decomposition of the
time-dependent wave function $\Psi_{tot}(x,t)$ with any given function $\mathcal{A}(k)$ into the sum of the time-dependent SWFs $\psi_{tr}(x,t)$
and $\psi_{ref}(x,t)$ (see (\ref{1f1}) and (\ref{2f1})). At the initial stage
\begin{eqnarray} \label{1t}
\psi_{tr,ref}(x,t)\simeq\psi_{tr,ref}^{in}(x,t)=\frac{1}{\sqrt{2\pi}} \int_{-\infty}^\infty \mathcal{A}(k) \Api_{tr,ref} e^{i(kx-E(k)t/\hbar)}dk;
\end{eqnarray}
At the final stage $\psi_{tr}(x,t)$ and $\psi_{ref}(x,t)$ approach $\Psi_{tr}(x,t)$ and $\Psi_{ref}(x,t)$, respectively.

Let $\textbf{T}=\langle\psi_{tr}|\psi_{tr}\rangle$ and $\textbf{R}=\langle\psi_{ref}|\psi_{ref}\rangle$. Considering (\ref{1t}) and (\ref{5}), it
is easy to show that at the initial and final stages
$$\textbf{T}=\int_{-\infty}^\infty
\mathcal{A}^2(k)T(k)dk\equiv\textbf{T}_{as},\ooo\textbf{R}=\int_{-\infty}^\infty \mathcal{A}^2(k)R(k)dk\equiv\textbf{R}_{as};$$
$\textbf{T}_{as}+\textbf{R}_{as}=1$. Note, at the initial stage this holds despite the fact that "transmission and reflection are inextricably
intertwined" because of interference between them. This takes place because (see (\ref{18}))
\begin{eqnarray*}
\langle\psi_{tr}^{in}|\psi_{ref}^{in}\rangle=\int_{-\infty}^\infty \mathcal{A}^2(k)\left[A^{in}_{tr}(k)\right]^*A^{in}_{ref}(k)dk =
i\int_{-\infty}^\infty \mathcal{A}^2(k) \sigma(k)\sqrt{T(k)R(k)}dk
\end{eqnarray*}
and, as a consequence, $\langle\psi_{tr}^{in}|\psi_{ref}^{in}\rangle+\langle\psi_{ref}^{in}|\psi_{tr}^{in}\rangle=0$.

Note, at the very stage of scattering, $\textbf{T}$ varies. Point is that the requirements (a)-(d) (Section \ref{separ}) ensure the balance of the
input $I_{tr}(x_c-0,k)$ and output $I_{tr}(x_c+0,k)$ probability flows only for single waves $\psi_{tr}(x,k)$ entering the wave packet
$\psi_{tr}(x,t)$. For the packet itself, the interaction between the main 'harmonic' $\psi_{tr}(x,k_0)$ and 'subharmonics' $\psi_{tr}(x,k)$ leads
to the imbalance between the input and output flows at the point $x_c$: $d\textbf{T}/dt=I_{tr}(x_c+0,t)-I_{tr}(x_c-0,t)\neq 0$. And, since the
role of subharmonics is essential at the leading and trailing fronts of the wave packet, this effect is maximal when these fronts cross the point
$x_c$. The total change of $\textbf{T}$, in the course of the OCS, is zero! As regards $\textbf{R}$, this norm remains constant even at the very
stage of scattering: $\textbf{R}(t)\equiv\textbf{R}_{as}$. This follows from the fact that $\psi_{ref}(x_c,t)=0$ and hence
$I_{ref}(x_c+0,t)=I_{ref}(x_c-0,t)=0$ for any value of $t$.

Of course, this effect, related to the transmission dynamics, is hidden behind the reflection subprocess. It can be observed only by means of
indirect measurements. As regards the OCS itself, the point $x_c$ is neither 'source' nor 'sink' for it -- this directly observable scattering
process is unitary.

\section{Discussion I: The cat paradox or how to reconcile, at the micro-level, the quantum-mechanical description of micro-cat states with
"classical physical theories"} \label{cat}

So, at the asymptotically large distances $\triangle x$ ($\triangle x\gg l_0$) from the potential barrier, transmission and reflection behave as
alternative subprocesses. At the same time, at the very stage of scattering, i.e., at the micro-scales, this is not. The transmission dynamics is
non-unitary at this stage.

Thus, we arrive at the following conclusion. At the macro-scales, the dynamics of the OCS's subprocesses respects QM and classical probability
theory. However, in the spatial interval which includes the barrier region and has the size of order  $l_0$, it violates both the 'unitary' QM and
probability theory. Of course, the second property was expected. It merely reflects the fact that the quantum probability differs in the general
case from its classical counterpart. The first property should be considered as a 'fee' for reconciling a quantum description of the OCS with "the
classical physical theories" at the macro-scales.

Note that the presented decomposition of the total wave function $\Psi_{tot}(x,t)$ into the SWFs $\psi_{tr}(x,t)$ and $\psi_{ref}(x,t)$, as well
as the prohibition of averaging over $\Psi_{tot}(x,t)$, is nothing but the reconciling of the quantum-mechanical description of this one-particle
micro-cat state with "classical physical theories", as it was required by Bohr \cite{Bohr}. The same should be done for any micro-cat state,
including the state of a radioactive atom in the Schr\"odinger's-cat thought-experiment. Thus, the first property of our model of the OCS can be
treated also as a 'fee' for solving the cat paradox at the micro-level.

But such a fee is quite justified. With this approach to micro-cat states, QM becomes compatible with the principles of a macroscopic realism
\cite{Leg} and, hence, micro-phenomena associated with micro-cat states become 'speakable'. At the same time, the existing approaches
\cite{Ghi,Sch,Joo} pay a higher price to solve this paradox. Whilst our approach discards only the current practice of modelling micro-cat states,
the approaches \cite{Ghi,Sch,Joo} discard the fundamental notion of a closed system, as well as discard, at the macro-level, the QM itself, with
its unitary dynamics (see, e.g., \cite{Leg}). In these approaches, "quantum mechanics looks as an ineffective theory, in which the micro-world is
"unspeakable" and macro-world is undescribable" \cite{Ch4}.

We have also to stress that Schr\"odinger himself \cite{Schr} did not consider his paradox as the 'macro-objectification' problem. For him it was
unacceptable to divide the world onto the micro-world described by QM and the macro-world governed by the laws of classical physics. The
proponents of this point of view were also Louis de Broglie and John Bell. Now this view is not widely accepted. Nevertheless, there are famous
physicists that support it. For example, Gerard 't Hooft says in \cite{Hooft} the following: "Many researchers are led to believe that the
microscopic world is controlled by 'a different kind of logic' than our classical logic. We insist that there exists only one kind of logic, even
if the observed phenomena are difficult to interpret".

Our model offers such a kind of logics. It not only opens a new way of solving the TTP since the transmission and reflection dynamics and their
peculiarities are known now at all stages of scattering. It also offers a cardinally new way of solving the problem of reconciling QM with
classical physics, according to which pure micro-cat states represent an intermediate link between elementary pure states and mixed states.

Our next step is to define the dwell and group times for subprocesses, as well as to consider the possibility of an indirect measurement of these
quantities by means of the Larmor-clock procedure. In doing so, we will pay a particular attention to the case of tunneling a particle through the
rectangular barrier in the opaque limit $d\to\infty$, when the group tunneling velocity, according to the existing approaches, must be
superluminal.

\section{Characteristic times for transmission and reflection} \label{f2}

\subsection{Dwell times} \label{f21}

Let us again, as in Section \ref{dwell}, apply the flow-velocity concept and the corresponding dwell-time concept for introducing the tunneling
(or, more generally, transmission) velocity and time in the stationary case. The transmission dwell time $\tau^{tr}_{dwell}$ reads as
\begin{eqnarray} \label{22}
\tau^{tr}_{dwell}(k)=\frac{1}{I_{tr}}\int_a^b|\psi_{tr}(x;k)|^2 dx.
\end{eqnarray}
As is seen, this time scale unlike the dwell times (\ref{8}) and (\ref{9}) is unambiguously associated with the transmission subprocess. Hence the
quantity $v_{flow}^{tr}(x)=I_{tr}/|\psi_{tr}(x;k)|^2$ can be surely treated as the average velocity of tunneling particles at the point $x$.

The reflection dwell time $\tau^{ref}_{dwell}$ is introduced as follows (see also \cite{But}):
$$\tau^{ref}_{dwell}(k)=\frac{1}{I_{ref}} \int_a^{x_c}|\psi_{ref}(x,k)|^2 dx,\ppp I_{ref}=\frac{\hbar
k}{m}R(k).$$ Note that $\tau^{ref}_{dwell}(k)$ depends not only on the average velocity of reflected particles, but also on the average depth of
their penetration into the barrier region. Therefore the expression $I_{ref}/|\psi_{ref}(x;k)|^2$ cannot be interpreted as the average velocity of
reflected particles at the point $x$.

It is useful here to compare the transmission dwell time $\tau^{tr}_{dwell}$ for a particle tunneling through the rectangular barrier of height
$V_0$ ($E<V_0$) (see \cite{Ch6,Ch2}) with the Buttiker dwell time $\tau_{dwell}^{(2)}$ \cite{But}:
\begin{eqnarray} \label{220}
\tau^{tr}_{dwell}=\frac{m}{2\hbar k\kappa^3}\left[\left(\kappa^2-k^2\right)\kappa d +\kappa_0^2 \sinh(\kappa d)\right],\ooo \tau_{dwell}^{(2)}=
\frac{m k}{\hbar \kappa}\cdot \frac{2\kappa d (\kappa^2-k^2)+\kappa_0^2 \sinh(2\kappa d)}{4k^2\kappa^2+ \kappa_0^4\sinh^2(\kappa d)};
\end{eqnarray}
$\kappa_0=\sqrt{2mV_0}/\hbar$.

As is seen, unlike $\tau_{dwell}^{(2)}$ the transmission dwell time $\tau^{tr}_{dwell}$ increases exponentially in the limit $d\to\infty$
($E<V_0$), rather than saturates. Thus, in our approach the dwell transmission time does not lead to the Hartman effect. The opaque barrier
strongly retards the motion of to-be-transmitted particles with a given energy $E$, when they enter into the barrier region.

\subsection{Asymptotic group times for transmission and reflection} \label{f22}

\begin{figure}[t]
\begin{center}
\includegraphics[scale=0.75]{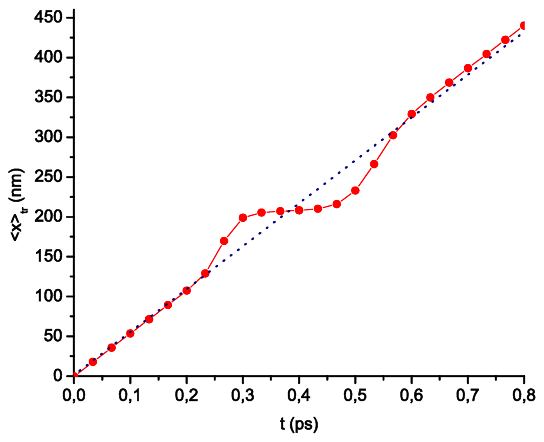}
\end{center}
\caption{The functions $X_{tr}(t)$ (solid line) and $X_{tr}^{in}(t)$ (dashed line)} \label{fig.5}
\end{figure}

Let us now consider the time-dependent transmission and reflection dynamics in the interval $[0,b+L]$, assuming that the incident wave packet
$\Psi_{tot}^{inc}$ is narrow in $k$ space ($l_0\gg d$). Let $X_{tr}(t)$ and $X_{ref}(t)$ be, respectively, the positions of the CMs of the wave
packets $\psi_{tr}(x,t)$ and $\psi_{ref}(x,t)$ at the instant $t$:
\begin{eqnarray*}
X_{tr}(t)=\frac{1}{\mathbf{T}}<\psi_{tr}|\hat{x}|\psi_{tr}>,\ppp X_{ref}(t)=\frac{1}{\mathbf{R}}<\psi_{ref}|\hat{x}|\psi_{ref}>.
\end{eqnarray*}
Fig.~\ref{fig.5} obtained for the 'opaque' rectangular barrier shows the results of tracing the CM's position of the wave packet $\psi_{tr}(x,t)$;
$a=200nm$, $b=215nm$, $V_0=0.2 eV$, $(\hbar k_0)^2/2m=0.05eV$ and $l_0=10nm$.

As is seen, at the asymptotically large distances from the barrier the velocity of the CM of $\psi_{tr}(x,t)$ is equal to $\hbar k_0/m$. However,
in the barrier region (see the almost flat part of the curve) the CM's velocity, like the flow velocity, is much smaller than $\hbar k_0/m$. That
is, the group-velocity concept justifies the effect of retardation of the to-be-transmitted wave packet in the barrier region, which was predicted
in the opaque limit on the basis of the flow-velocity concept. Note that in the case under consideration the wave packet $\psi_{tr}(x,t)$ is much
wider than the barrier. So that, when the CM of this packet is moving within the barrier region, its front and tail fronts are moving outside this
region. In this case, the main harmonic $\psi_{tr}(x;k_0)$ dominates in the barrier region and, thus, its interaction at the midpoint $x_c$ with
subharmonics is negligible. As a consequence, at this stage, this point does not influence the norm of the wave packet $\psi_{tr}(x,t)$ and the
velocity of its CM (see also Section \ref{separ}).

However, its influence is essential when the front or tail part of this narrow (in $k$ space) wave packet crosses this point. Namely, when its
{\it front} part crosses the midpoint of the region of the {\it opaque} rectangular barrier, this point serves as a (hidden) source of particles.
This results in accelerating the CM of the wave packet, what has nothing to do with the average velocity of transmitted particles. On the
contrary, when the tail part of the packet crosses this point, then this point serves as a 'sink' for this subensemble. And, since the main body
of the packet is located at this stage to the right of the point, this effect leads again to accelerating the CM of the packet (what, again, has
nothing to do with the average velocity of transmitted particles). As a result, the group time to describe the CM's transmission dynamics in the
interval $[0,b+L]$ proves to be anomalously short in the opaque limit.

The transmission and reflection group times for this asymptotically large interval can be defined as follows. Let $t^{tr}_{depart}$ and
$t^{tr}_{arrive}$ be such instants of time that
\begin{eqnarray} \label{23}
X_{tr}(t^{tr}_{depart})=0;\ppp X_{tr}(t^{tr}_{arrive}) =b+L,
\end{eqnarray}
Then the transmission time $\Delta t_{tr}$ for this interval can be defined as the difference $t^{tr}_{arrive}- t^{tr}_{depart}$.

Similarly, let $t^{ref}_{depart}$ and $t^{ref}_{arrive}$ be such instants of time that
\begin{eqnarray} \label{24}
X_{ref}(t^{ref}_{depart})=X_{ref}(t^{ref}_{arrive}) =0;
\end{eqnarray}
$t^{ref}_{depart}$ and $t^{ref}_{arrive}$ are, respectively, the smallest and largest roots of Eq. (\ref{24}); since $a\gg l_0$ these roots are
evident to exist. Then the reflection time $\Delta t_{ref}$ can be defined as follows: $\Delta t_{ref}=t^{ref}_{arrive}-t^{ref}_{depart}$.

Notice, since all quantities in (\ref{23}) and (\ref{24}) are associated with the asymptotically remote spatial points, we can calculate them
through the incoming and outgoing wave packets (\ref{5}). For narrow packets the positions $X_{tr}(t)$ and $X_{ref}(t)$ at the initial and final
stages of scattering are defined as follows (the CM position $X_{tot}(t)$ for the total wave packet at the initial stage of scattering is
presented here too):

(a) long before the scattering event $$ X_{tr}(t)=X_{ref}(t)\simeq\frac{\hbar k}{m}t-\sigma(k)\lambda'(k),\ppp X_{tot}(t)\simeq
X_{tot}^{in}(t)=\frac{\hbar k}{m}t;$$

(b) long after the scattering event $$X_{tr}(t)\simeq X_{tr}^{out}=\frac{\hbar k}{m}t-J'(k)+d,\ppp X_{ref}(t)\simeq X_{ref}^{out}= -\frac{\hbar
k}{m}t-J'(k)+2a.$$

Thus, taking into account these expressions in (\ref{23}) and (\ref{24}) and substituting $k_0$ by $k$, we obtain
\begin{eqnarray*}
\Delta t_{tr}=\frac{m}{\hbar k}\left[J'(k)-\sigma(k)\lambda'(k)+a+L\right],\ooo \Delta t_{tr}=\Delta t_{ref}(L)=\frac{m}{\hbar
k}\left[J'(k)-\sigma(k)\lambda'(k)+2a\right].
\end{eqnarray*}

And lastly, excluding from these expressions the terms to describe the outer spatial regions, we obtain the asymptotic scattering times
$\tau^{tr}_{as}$ and $\tau^{ref}_{as}$ for transmission and reflection, respectively:
\begin{eqnarray} \label{127}
\tau^{tr}_{as}=\tau^{ref}_{as}=\frac{m}{\hbar k}\left[J'(k)-\sigma(k)\lambda'(k)\right].
\end{eqnarray}
The corresponding delay times $\tau^{tr}_{del}=\tau^{tr}_{as}-\tau_{free}$ and $\tau^{ref}_{del}=\tau^{ref}_{as}-\tau_{free}$, where
$\tau_{free}=md/\hbar k$, are
\begin{eqnarray} \label{128}
\tau^{tr}_{del}=\tau^{ref}_{del}=\frac{m}{\hbar k}\left[J'(k)-d-\sigma(k)\lambda'(k)\right].
\end{eqnarray}
As is seen, unlike Exp. (\ref{10}) for the corresponding time scales in the conventional model of the OCS, Exps. (\ref{127}) and (\ref{128})
contain the extra term with $\lambda'(k)$.

For tunneling trough the rectangular barrier ($E<V_0$), the asymptotic transmission time $\tau^{tr}_{as}$ and the starting position
$X_{start}=X_{tr}(0)=X_{ref}(0)$ are defined by the expressions (see \cite{Ch2})
\begin{eqnarray} \label{129}
\tau^{tr}_{as}=\frac{4m}{\hbar k\kappa}\ooa\frac{[k^2+\kappa_0^2 \sinh^2(\kappa d/2)][\kappa_0^2 \sinh(\kappa d)-k^2\kappa d]}{4k^2\kappa^2+
\kappa_0^4\sinh^2(\kappa d)},\\
X_{start}=-2\frac{\kappa_0^2}{\kappa}\ooa \frac{(\kappa^2-k^2)\sinh(\kappa d)+k^2\kappa d \cosh(\kappa d)}{4k^2\kappa^2+ \kappa_0^4\sinh^2(\kappa
d)}.\nonumber
\end{eqnarray}
From Exps.~(\ref{129}) it follows that $\tau^{tr}_{as}\to 2m/(\hbar k\kappa)$ and $X_{subpr}^{in}(0)\to X_{tot}^{in}(0)=0$ in the limit
$d\to\infty$. These two results agree with Fig.~\ref{fig.5} to show that, for tunneling through the opaque barrier, the curve $X_{tr}(t)$ indeed
evolves from the origin and the group transmission time $\tau^{tr}_{as}$ are much smaller than the dwell transmission time.

For example, in the case to correspond to Fig.~\ref{fig.5}, $\tau^{tr}_{as}\approx 0.024ps$ and $\tau^{tr}_{dwell}\approx 0.155ps$. In this case,
$\tau^{tr}_{as}$ describes the influence of the barrier on the subensemble of transmitted particles in the asymptotically large interval $0,b+L$,
while $\tau^{tr}_{dwell}$ describes the transmission dynamics in the barrier region.

\begin{figure}[t]
\begin{center}
\includegraphics[scale=0.7]{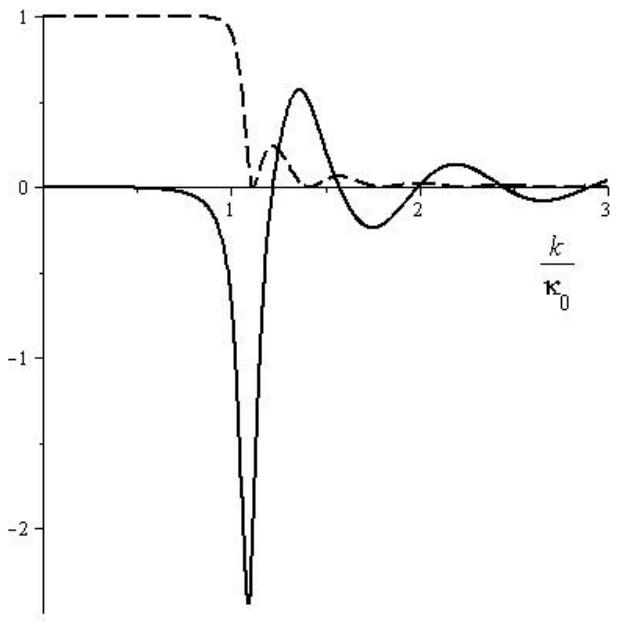}
\end{center}
\caption{$X_{start}/d$ (full line) and the reflection coefficient $R$ (broken line) as functions of $k/\kappa_0$; $\kappa_0 d =2\pi $.}
\label{x_start}
\end{figure}

As is seen from Exp. (\ref{129}), $X_{start}\to X_{tot}^{in}(0)=0$ in the opaque limit. That is, in fact, our approach confirms the existence of
the Hartman effect predicted in this limit on the basis of the Wigner tunneling time $\tau_W$ (see Section \ref{hart1})! But this fact does not
all mean that our approach justifies the concept of the Wigner tunneling time. As it follows from (\ref{129}), $X_{start}\neq 0$ in the general
case (see Fig.~\ref{x_start}). The difference between $\tau^{tr}_{as}$ and $\tau_W$ is maximal at the resonance points where $R=0$. However, the
most dramatic difference between these time scales occurs in the long-wave limit $k\to 0$:
\begin{eqnarray} \label{E0}
X_{start}=-\frac{2}{\kappa_0\sinh(\kappa_0 d)},\ppp \frac{\tau^{tr}_{as}}{\tau_{free}}=\frac{2}{\kappa_0 d}\tanh\left(\frac{\kappa_0 d}{2}\right);
\end{eqnarray}
while
\begin{eqnarray} \label{E1}
X_{tot}^{in}(0)=0,\ppp \frac{\tau_{free}}{\tau_W}=\frac{\kappa_0 d}{2} \tanh\left(\frac{\kappa_0 d}{2}\right).
\end{eqnarray}
As is seen from Exp. (\ref{E1}), in the limit $k\to 0$ and $d\to 0$ the Wigner tunnelling time diverges, while the tunneling time (\ref{E0})
approaches $\tau_{free}$.

Our next step is to consider the Larmor-clock procedure \cite{But} adapted in \cite{Ch2,Ch9} to the transmission and reflection subprocesses.

\subsection{Larmor times for transmission and reflection} \label{f4}

For the rectangular barrier, the wave functions $\psi^{(\uparrow)}(x;k)$ and $\psi^{(\downarrow)}(x;k)$, that describe, respectively, the
stationary states of particles with spins parallel and antiparallel to the external infinitesimally small magnetic field, switched on within the
barrier region, can be written as follows,
\begin{eqnarray} \label{27}
\psi_{tr,ref}^{(\uparrow,\downarrow)}(x;k)=\psi_{tr,ref}(x;k)\mp \omega_L \tilde{\psi}_{tr,ref}(x;k),\ppp
\tilde{\psi}_{tr,ref}(x;k)=\frac{m}{2\hbar \kappa}\frac{\partial \psi_{tr,ref}(x;k)}{\partial \kappa};
\end{eqnarray}
here $\omega_L=2\mu B/\hbar$, $\mu$ and $B$ are, respectively, the absolute values of the electron magnetic moment and applied magnetic field. As
was shown in \cite{Ch2,Ch9} for the limiting case $l_0\to\infty$, for the rotation angles $\Delta \phi^{tr,ref}_\bot$ and $\Delta
\phi^{tr,ref}_\|$, which describe the Larmor precession of the average spins of transmitted and reflected particles in the plane orthogonal to the
magnetic field and, respectively, their "aligning" with the field, we have
\begin{eqnarray} \label{28}
\Delta \phi^{tr}_\bot\equiv -\omega_L (\tau_\bot^{tr}-\tau_{0\bot}^{tr})=-\omega_L (\tau^{tr}_{dwell}+\tau_{flip}),\ooo \Delta \phi^{tr}_\|\equiv
-\omega_L (\tau_\|^{tr}-\tau_{0\|}^{tr})=0,\\
\Delta \phi^{ref}_\bot\equiv -\omega_L (\tau_\bot^{ref}-\tau_{0\bot}^{ref})=-\omega_L \tau^{tr}_{dwell},\ooo \Delta \phi^{ref}_\|\equiv -\omega_L
(\tau_\|^{ref}-\tau_{0\|}^{ref})=0;\nonumber
\end{eqnarray}
here $\tau_{\bot,\|}^{tr,ref}$ and $\tau_{0\bot,\|}^{tr,ref}$ are the final and initial readings of the Larmor clocks, that obey the relationships
\begin{eqnarray} \label{29}
\tau_\bot^{tr}=\tau_\bot^{ref}\equiv\tau_\bot,\ooo \tau_{0\bot}^{tr}=\tau_{0\bot}^{ref}\equiv\tau_{0\bot}\neq 0,\ooo
T\tau_{0\|}^{tr}+R\tau_{0\|}^{ref}=0;\ooo \tau_{0,\bot}^{tr}=\tau_{0\|}^{tr}\sqrt{T/R}\ooo (E<V_0);
\end{eqnarray}
$\tau^{tr}_{dwell}$ and $\tau^{ref}_{dwell}$ define the duration of the Larmor precession; $\tau_{flip}$ is defined by the expression
\begin{eqnarray} \label{30}
\tau_{flip}(k)=\frac{1}{kT(k)}\Re\Bigg[\psi_{tr}(x_c;k)\left(\frac{\partial\tilde{\psi}_{tr}^*(x_c+0;k)}{\partial
x}-\frac{\partial\tilde{\psi}_{tr}^*(x_c-0;k)}{\partial x}\right)\nonumber\\
-\tilde{\psi}_{tr}^*(x_c;k)\left(\frac{\partial\psi_{tr}(x_c+0;k)}{\partial x}-\frac{\partial\psi_{tr}(x_c-0;k)}{\partial x}\right)\Bigg].
\end{eqnarray}
With taking into account the relationships (\ref{180}) and (\ref{27}), this quantity can be rewritten as
\begin{eqnarray} \label{31}
\tau_{flip}(k)=\frac{m}{\hbar k\kappa T(k)}\left(M_{tot}\frac{\partial^2 M_{tot}}{\partial x \partial\kappa}-\frac{\partial M_{tot}}{\partial x}
\frac{\partial M_{tot}}{\partial \kappa}\right)\Bigg|_{x=x_c};
\end{eqnarray}
here $M_{tot}=|\Psi_{tot}(x;k)|$.

Note, according to the Larmor-clock procedure \cite{But} (see Section \ref{larmor})
\begin{eqnarray*}
\tau_{0\bot}^{tr}=\tau_{0\bot}^{ref}=\tau_{0\|}^{tr}=\tau_{0\|}^{ref}=0,\ooo \Delta \phi^{tr}_\|= -\omega_L \tau_\|^{tr}\neq 0,\ooo \Delta
\phi^{ref}_\|= -\omega_L \tau_\|^{ref}\neq 0.
\end{eqnarray*}
Nonzero values of $\Delta \phi^{tr}_\|$ and $\Delta \phi^{ref}_\|$ say that this approach allows aligning the (average) electron's spin with the
magnetic field, what contradicts QM. In our approach this effect is absent: $\tau_\|^{tr}=\tau_{0\|}^{tr}$ and $\tau_\|^{ref}=\tau_{0\|}^{ref}$.

Another important difference consists in that in our Larmor-clock model there are {\it two} physical effects that influence the average spin of
transmitted particles in the plane orthogonal to the magnetic field, rather than one as in the standard approach. Apart from the already known
Larmor precession of this spin under the magnetic field, whose duration is described by the dwell time $\tau^{tr}_{dwell}$, there appears a new
effect -- flipping the orthogonal projection of the average spin at the point $x_c$ --  which is described by the quantity $\tau_{flip}$. This
effect does not allow a direct measurement of $\tau^{tr}_{dwell}$. For reflected particles this effect does not appear because
$\psi_{ref}(x_c,t)=0$ (see (\ref{17})); the substitution $\psi_{ref}$ for $\psi_{tr}$ in (\ref{30}) yields zero value.

As it follows from Eqs. (\ref{27}) and (\ref{28}), for both subprocesses in the case $E<V_0$ we have
\begin{eqnarray} \label{33}
\tau^{tr}_{dwell}+\tau_{flip}=\tau^{ref}_{dwell}=\tau_\bot-\tau_{0\bot}=\tau_\bot+\tau^{ref}_\| \sqrt{R/T}=\tau_\bot-\tau^{tr}_\| \sqrt{T/R}
\end{eqnarray}
where $T=4k^2\kappa^2/[4k^2\kappa^2+\kappa_0^4\sinh^2(\kappa d)]$;
\begin{eqnarray*}
\tau_\bot=\frac{mk}{\hbar\kappa}\ooa \frac{2(\kappa^2-k^2)\kappa d+\kappa_0^2\sinh(2\kappa d)}{4k^2\kappa^2+ \kappa_0^4\sinh^2(\kappa d)},\ooo
\tau_{0\bot}=\frac{2mk}{\hbar\kappa}\ooa \frac{(\kappa^2-k^2)\sinh(\kappa d)+\kappa_0^2\kappa d \cosh(\kappa d)}{4k^2\kappa^2+
\kappa_0^4\sinh^2(\kappa d)}.
\end{eqnarray*}
Note, for $E>V_0$ the relationship between $\tau_{0,\bot}^{tr}$ and $\tau_{0\|}^{tr}$ (see (\ref{29})) is slightly different.

Thus, in the opaque limit, when the transmission dwell time $\tau^{tr}_{dwell}$ increases exponentially, the final readings $\tau_\bot$ of the
Larmor clock saturate together with the reflection dwell time $\tau^{ref}_{dwell}$. In this case the time $\tau_{flip}$ is negative and its
absolute value increases exponentially. The quantity $\tau_{0\bot}$ to describe the initial readings of the Larmor clock for both subprocesses
tends to zero in this case; its analog in the timekeeping procedure based on the group-velocity concept, $\tau_{group}^{(0)}=-m X_{start}/\hbar k$
(see (\ref{129})), diminishes in this limit too.

\section{Discussion II: On the Hartman effect predicted on the basis of $\tau^{tr}_{as}$} \label{hart1}

So, for narrow wave packets scattering at the rectangular barrier in the opaque limit $d\to \infty$ ($E<V_0$), the asymptotic group transmission
time $\tau^{tr}_{as}$ saturates like the Wigner time $\tau_W$. In this limit, $\tau^{tr}_{as}$ coincides with $\tau_W$ because the initial
position $X_{start}$ of the CM of the to-be-transmitted wave packet approaches $X^{in}_{tot}(0)=0$ (of course, $\psi_{tr}(x,t)$ does not approach
$\Psi_{tot}(x,t)$ in this case). Similarly, in the Larmor-clock procedure the difference $\tau_\bot-\tau_{0\bot}$ saturates and $\tau_{0\bot}\to
0$, in this limit.

In this sense, our concept $\tau^{tr}_{as}$ confirms the existence of the Hartman effect. However, unlike the conventional approach ours says that
this time scales has nothing to do with the average velocity of tunneling {\it particles} in the barrier region. Only the dwell time $\tau_{tr}$
and local group time (see fig.~\ref{fig.5}) are associated with this velocity. As regards $\tau^{tr}_{as}$ and $\tau_\bot$ based, respectively, on
tracing the {\it average position} of the subensemble of transmitted particles and their {\it average spin}, both depend on extra effects which
have no relation to the average velocity of tunneling {\it particles} in the barrier region.

As was shown in Section \ref{f22}, $\tau^{tr}_{as}$ characterizes the transmission dynamics at the asymptotically large spatial region $[0,b+L]$.
At the stages, when the front and tail parts of the narrow wave packet $\psi_{tr}(x,t)$ are crossing the point $x_c$, the velocity of its CM
varies because of changing 'the number of particles' at the point $x_c$ in the subensemble described by this packet. In the opaque limit this
effect leads to the speed-up of the CM of the packet (see Fig.~\ref{fig.5}) and, in the last analysis, to superluminal group tunneling velocities.
But this nonlocal speed-up of the wave packet $\psi_{tr}(x,t)$ is associated with the nonunitarity of the transmission dynamics at these two
(sub)stages of the very stage of scattering. This speed-up mechanism makes it impossible to directly observr the low velocity of the
$\psi_{tr}(x,t)$'s CM in the barrier region! (To some extent, this tunneling scenario resembles the 'intuitive' scheme with virtual photons, which
was proposed by Nimtz (see Section \ref{Nimtz}.)

As regards the Larmor-clock procedure, again, due to another effect -- "flipping" the average spin of transmitted particles at the point $x_c$, it
does not allow one to directly measure the dwell time $\tau^{tr}_{dwell}$. In the opaque limit, this effect leads to the situation when the final
readings of the Larmor clock show the time that is much smaller than the average time spent by tunneling particles in the barrier region. The time
$\tau^{tr}_{dwell}$ spent by transmitted particles in the barrier region can be measured indirectly with making use of the relationships
(\ref{28}), (\ref{29}) and (\ref{33}) between characteristic times of both subprocesses (see also Exp. (\ref{31}) for $\tau_{flip}(x;k)$).

\section{Conclusion}\label{conclude}

We show that the TTP is not an ill-posed problem. Rather, the TTP is a problem which is in principle insolvable within the framework of the
conventional model of the OCS. Contrary to Bohr's requirement (see Section \ref{req}), this model not only is incompatible with the "classical
physical theories", but also is {\it internally} inconsistent (see Sections \ref{dwell} and \ref{Bohm}).

We argue that the TTP stands alongside with such fundamental problem of QM as the Schr\"odinger's-cat and EPR-Bohm paradoxes. Moreover, the TTP
and cat paradox have the same root: resolving these two quantum-mechanical problems requires revising the current practice of description of pure
micro-cat states. This practice is based on the Bell theory of the EPR-Bohm thought-experiment, according to which LHVs do not exist, as well as
on the interpretation of micro-cat states, according to which a particle in a micro-cat state occupies simultaneously two macroscopically distinct
(sub)states that form this micro-cat state. It is this practice that makes it impossible a correct resolving of the TTP.

At the same time, according to the probabilistic approach \cite{Acc1,Acc,Khr1,Khr2,Phil,Hess}, the experimental violation of Bell's inequality
does not at all mean the non-existence of LHVs. This experimental fact means, rather, that the probability distributions to enter into this
inequality describe mutually incompatible statistical data. That is, in fact, Bell's theory of the EPR-Bohm thought-experiment, developed on the
basis of classical probability theory, contradicts this theory. Bell derived his famous inequality for {\it non-Kolmogorovian} LHVs and hence
there is nothing surprising in the fact that this inequality is violated in experiment.

We argue that the Kolmogovness requirement should be extended also onto the quantum-mechanical state of a scattered particle, in the case of the
OCS, as well as onto the state of a radioactive atom in the cat paradox. Indeed, in both these cases we deal with the so-called micro-cat states.
The squared modulus of either micro-cat state describes mutually incompatible statistical data and hence its interpretation as the probability
density contradicts probability theory.

According to our program of solving the TTP and cat paradox, the adequate quantum-mechanical description of the OCS and Schr\"odinger's
thought-experiment, compatible with "classical physical theories", must represent the time-dependent state of each micro-object (a particle and
radioactive atom) as a micro-cat state (i.e., as the superposition of two alternative sub-states) for any value of $t$. In this case we assume,
following Volovich \cite{Vol1,Vol2}, that the quantum dynamics of a scattering particle and a radioactive atom (and hence the time-dependent state
of each micro-object) must obey the necessary condition: it "must be embedded into the spacetime structure, being correct from the viewpoint of
relativity theory". Quantum models violating this requirement should be considered as purely speculative models which create a poor basis for
judging about a quantum locality (or non-locality) of micro-cat states.

We have to stress that making use, in Schr\"odinger's experiment, of such abstract states as 'decayed atom' and 'undecayed atom', that 'live'
beyond space and time, is inadmissible in studying the quantum {\it dynamics} of decaying a radioactive atom. As is known, in the simplest
quantum-mechanical model of this phenomenon, it can be described as a tunneling of a less massive fragment of the atom from the potential well
created by its more massive fragment. Thus, it is more correct to substitute the macroscopically distinct states -- the 'bound state' (when the
lightest fragment is still situated in the well) and the 'unbound state' (when it has already tunneled from the well) -- for the abstract states
'non-decayed atom' and 'decayed atom', respectively. The main peculiarity of the first pair of states is that, unlike the last pair, they must be
"embedded into the spacetime structure" by definition.

According to our approach, in the quantum-mechanical model of decaying the radioactive atom, compatible with "classical physical theories", the
sum of the 'bound state' and 'unbound state' should yield the (micro-cat) state of the radioactive atom, and the sum of their norms should yield
the norm of this micro-cat state. And only the squared modula of 'bound state' and 'unbound state' have the meaning of the probability densities
-- the cat paradox must be solved at the micro-level. According to the presented approach, its solving reduces to the problem of decomposition of
the original state of a radioactive atom into the sum of 'bound state' and 'unbound state' of its more easy piece.

Note that Bell's classical-like description of the EPR-Bohm experiment violates, in fact, not only classical probability theory, but also the
requirement \cite{Vol1,Vol2}, because the quantum dynamics of the electron EPR-pair, in this description, has not been "embedded into a correct
spacetime structure". That is, the current theories of this famous experiment must be revised. And what is important is that this concerns not
only its Bell's classical-like theory, but also its modern quantum model.

The presented 'reconciliation' programm has been performed in this paper by the example of the OCS. For symmetric potential barriers we have
developed a 'probabilistic' model of the OCS, where we represented $\Psi_{tot}$ as a micro-cat state, for all stages of scattering. Our approach
allows one to uniquely reconstruct the whole prehistory of each subprocess according to their final states $\Psi_{tr}$ and $\Psi_{ref}$. At the
initial and final stages of scattering, i.e., at the asymptotically large distances from the barrier, the quantum probability in this model agrees
with its classical counterpart. At the very stage of scattering the quantum probabilities, describing the dynamics of both subprocesses, behave
non-classically.

Note that the transmission dynamics is not unitary at this stage, and at first glance this property contradicts QM. But this is not, because the
formalism of decomposing the original (unitary) quantum scattering process into alternative subprocesses is beyond the conventional
quantum-mechanical practice. Figuratively speaking, the non-unitarity of the transmission dynamics at the very stage of scattering should be
considered as a 'fee' for solving the cat paradox at the micro level.

On the basis of the subprocesses' wave functions we have defined characteristic times for each subprocess. Our concept of the asymptotic group
transmission time confirms the existence of the Hartman effect predicted on the basis of the Wigner group time in the opaque limit. We show that
this effect has nothing to do with the (average) velocity of tunneling {\it particles}. The latter can be derived only from the dwell transmission
time which grows exponentially in the opaque limit. In the general case our model does not support the Wigner-time concept, irrespective of the
fact whether the corresponding tunneling velocity is superluminal or subluminal. All characteristic times for transmission and reflection can be
measured only indirectly because both the subprocesses hide each other. Such measuring can be performed, for example, by means of the Larmor-clock
time-keeping procedure (see Section \ref{f4}) based on the presented model of the OCS.

It is also worthwhile to note that, in the case of the OCS, our approach implies the introduction, on the basis of the wave functions $\psi_{tr}$
and $\psi_{ref}$, of two sets of Bohmian trajectories. Now, each point in the region $x<a$ serves as the starting point of two trajectories,
rather than one: one goes to plus infinity, while the other goes to minus infinity. Within our approach, the conventional (totally nonlocal)
Bohmian model of the OCS is transformed into the model, local at the scales much more the wave-packet's width. And what is also important, though
each single Bohmian trajectory of a particle taking part in the OCS does not coincide with the corresponding classical one (see, e.g.,
\cite{Hol}), the ensemble of Bohmian trajectories is equivalent to that of classical trajectories (see \cite{Ch10}).

To some extent the Bohmian model is equivalent to 'hydrodynamical' approaches to quantum processes (see, e.g., \cite{Hof} and references therein).
In the last analysis, all such theories, together with QM, model the dynamics of a micro-particle at the level of ensembles. As regards modelling
the dynamics of single members of quantum one-particle ensembles, this is the destiny of a future sub-quantum theory. Currently, research in this
direction are already underway (see, e.g., \cite{Hest,Khr7,Hooft1}).

\section*{Acknowledgments}

This work has been partially financed by the Programm of supporting the leading scientific schools of RF (grant No 88.2014.2).


\end{document}